\documentclass[journal]{aiaa-pretty}
\usepackage{cite}

\newcommand{\pos}{p} 
\newcommand{\Pos}{\bar{p}} 
\newcommand{\vel}{v} 
\newcommand{\Vel}{\bar{v}} 
\newcommand{\hw}{\mathbb{H}} 
\newcommand{\hws}{\mathbb{S}} 
\newcommand{\hwd}{\hat{d}}
\newcommand{\cost}[1]{c_\text{#1}}
\newcommand{\wpt}{\mathcal{W}}
\newcommand{\cmap}{c}
\newcommand{\ccost}{C}
\newcommand{\ppath}{\mathbb{P}}
\newcommand{\ocost}{V}

\newcommand{\sepdist}{d_\text{sep}} 

\newcommand{\td}{t_\text{faulty}} 
\newcommand{\veh}[1]{Q_{#1}}
\newcommand{\vehSCS}[1]{\mathcal{Q}_{#1}} 

\author{Mo Chen\thanks{PhD Candidate, Department of Electrical Engineering and Computer Sciences}, Qie Hu\thanksibid{1}, Jaime F. Fisac\thanksibid{1}, Kene Akametalu\thanksibid{1}, Casey Mackin\thanks{PhD Student, Department of Electrical Engineering and Computer Sciences}, Claire J. Tomlin\thanks{Professor, Department of Electrical Engineering and Computer Sciences, Member AIAA}\\\textit{University of California, Berkeley}}
\title{Reachability-Based Safety and Goal Satisfaction of Unmanned Aerial Platoons on Air Highways}

\AIAAabstract{Recently, there has been immense interest in using unmanned aerial vehicles (UAVs) for civilian operations. As a result, unmanned aerial systems traffic management is needed to ensure the safety and goal satisfaction of potentially thousands of UAVs flying simultaneously. Currently, the analysis of large multi-agent systems cannot tractably provide these guarantees if the agents' set of maneuvers is unrestricted. In this paper, platoons of UAVs flying on air highways is proposed to impose an airspace structure that allows for tractable analysis. For the air highway placement problem, the fast marching method is used to produce a sequence of air highways that minimizes the cost of flying from an origin to any destination. The placement of air highways can be updated in real-time to accommodate sudden airspace changes. Within platoons traveling on air highways, each vehicle is modeled as a hybrid system. Using Hamilton-Jacobi reachability, safety and goal satisfaction are guaranteed for all mode transitions. For a single altitude range, the proposed approach guarantees safety for one safety breach per vehicle; in the unlikely event of multiple safety breaches, safety can be guaranteed over multiple altitude ranges. We demonstrate the platooning concept through simulations of three representative scenarios.}

\begin{document}
\maketitle
\section*{Nomenclature}
\noindent\begin{tabular}{@{}lcl@{}}
$\cmap$ &=& Cost map \\
$\ppath$ &=& A path between two points \\
$\ccost$ &=& Cumulative cost of a path \\
$\ocost$ &=& Value function of partial differential equations \\
$\hw$ &=& Air highway \\
$\hwd$ &=& Direction of travel of air highway \\
$\hws$ &=& A sequence of air highways \\
$\wpt$ &=& Waypoint \\
$x$ &=& System state (of a vehicle) \\
$\pos=(\pos_x, \pos_y)$ &=& Horizontal position \\
$\vel=(\vel_x, \vel_y)$ &=& Horizontal velocity \\
$\Pos$ &=& Target position \\
$\Vel$ &=& Target velocity \\
$\sepdist$ &=& Separation distance of vehicles within a platoon \\
$\td$ &=& Time limit for descent during potential conflict \\
$\veh{i}$ &=& $i$th vehicle \\
$\vehSCS{i}$ &=& Set of vehicles for vehicle $\veh{i}$ to consider for safety 
\end{tabular} \\

\section{Introduction}
Unmanned aerial vehicles (UAVs) have in the past been mainly used for military operations \cite{Tice91, Haulman2003}; however, recently there has been an immense surge of interest in using UAVs for civil applications. Through projects such as Amazon Prime Air \cite{PrimeAir} and Google Project Wing \cite{ProjectWing}, companies are looking to send UAVs into the airspace to not only deliver commercial packages, but also for important tasks such as aerial surveillance, emergency supply delivery, videography, and search and rescue \cite{Kopardekar16}. In the future, the use of UAVs is likely to become more and more prevalent.

As a rough estimate, suppose in a city of 2 million people, each person requests a drone delivery every 2 months on average and each delivery requires a 30-minute trip for a UAV. This would equate to thousands of UAVs simultaneously in the air just from package delivery services. Applications of UAVs extend beyond package delivery; they can also be used, for example, to provide supplies or to respond to disasters in areas that are difficult to reach but require prompt response \cite{Debusk10,Tornado16}. As a result, government agencies such as the Federal Aviation Administration (FAA) and National Aeronautics and Space Administration (NASA) are also investigating unmanned aerial systems (UAS) traffic management (UTM) in order to prevent collisions among potentially numerous UAVs \cite{Kopardekar16, FAA13, NASA16}. 

Optimal control and game theory present powerful tools for providing safety and goal satisfaction guarantees to controlled dynamical systems under bounded disturbances, and various formulations \cite{Bokanowski10, Mitchell05, Barron89} have been successfully used to analyze problems involving small numbers of vehicles \cite{Fisac15, Chen14, Chen17, Ding08}. These formulations are based on Hamilton-Jacobi (HJ) reachability, which can compute the backward reachable set (BRS), defined as the set of states from which a system is guaranteed to have a control strategy to reach a target set of states. HJ reachability is a powerful tool because BRS can be used for synthesizing both controllers that steer the system away from a set of unsafe states (``safety controllers'') to guarantee safety, and controllers that steer the system into a set of goal states (``goal satisfaction controllers'') to guarantee goal satisfaction. Unlike many formulations of reachability, the HJ formulations are flexible in terms of system dynamics, enabling the analysis of controlled nonlinear systems under disturbances. Furthermore, HJ reachability analysis is complemented by many numerical tools readily available to solve the associated HJ partial differential equation (PDE) \cite{LSToolbox, Osher02, Sethian96}. However, the computation is done on a grid, making the problem complexity scale exponentially with the number of states, and therefore with the number of vehicles. Consequently, HJ reachability computations are intractable for large numbers of vehicles. 

In order to accommodate potentially thousands of vehicles simultaneously flying in the air, additional structure is needed to allow for tractable analysis and intuitive monitoring by human beings. An air highway system on which platoons of vehicles travel accomplishes both goals. However, many details of such a concept need to be addressed. Due to the flexibility of placing air highways compared to building ground highways in terms of highway location, even the problem of air highway placement can be a daunting task. To address this, in the first part of this paper, we propose a flexible and computationally efficient method based on \cite{Sethian96} to perform optimal air highway placement given an arbitrary cost map that captures the desirability of having UAVs fly over any geographical location. We demonstrate our method using the San Francisco Bay Area as an example. Once air highways are in place, platoons of UAVs can then fly in fixed formations along the highway to get from origin to destination. The air highway structure greatly simplifies safety analysis, while at the same time allows intuitive human participation in unmanned airspace management.

A considerable body of work has been done on the platooning of ground vehicles \cite{Kavathekar11}. For example, \cite{McMahon90} investigated the feasibility of vehicle platooning in terms of tracking errors in the presence of disturbances, taking into account complex nonlinear dynamics of each vehicle. \cite{Hedrick92} explored several control techniques for performing various platoon maneuvers such as lane changes, merge procedures, and split procedures. In \cite{Lygeros98}, the authors modeled vehicles in platoons as hybrid systems, synthesized safety controllers, and analyzed throughput. Reachability analysis was used in \cite{Alam11} to analyze a platoon of two trucks in order to minimize drag by minimizing the following distance while maintaining collision avoidance safety guarantees. Finally, \cite{Sabau16} provided a method for guaranteeing string stability and eliminating accordion effects for a heterogeneous platoon of vehicles with linear time-invariant dynamics.

Previous analyses of a large number of vehicles typically do not provide safety and goal satisfaction guarantees to the extent that HJ reachability does; however, HJ reachability typically cannot be used to tractably analyze a large number of vehicles. In the second part of this paper, we propose organizing UAVs into platoons, which provides a structure that allows pairwise safety guarantees from HJ reachability to better translate to safety guarantees for the whole platoon. With respect to platooning, we first propose a hybrid systems model of UAVs in platoons to establish the modes of operation needed for our platooning concept. Then, we show how reachability-based controllers can be synthesized to enable UAVs to successfully perform mode switching, as well as prevent dangerous configurations such as collisions. Finally, we show several simulations to illustrate the behavior of UAVs in various scenarios.

Overall, this paper is not meant to provide an exhaustive solution to the unmanned airspace management problem. Instead, this paper illustrates that the computation intractability of HJ reachability can be overcome using an air highway structure with UAVs flying in platoons. In addition, the results are intuitive, which can facilitate human participation in managing the airspace. Although many challenges not addressed in this paper still need to be overcome, this paper can provide a starting point for future research in large-scale UASs with safety and goal satisfaction guarantees.
\section{Air Highways}
We consider air highways to be virtual highways in the airspace on which a number of UAV platoons may be present. UAVs seek to arrive at some desired destination starting from their origin by traveling along a sequence of air highways. Air highways are intended to be the common pathways for many UAV platoons, whose members may have different origins and destinations. By routing platoons of UAVs onto a few common pathways, the airspace becomes more tractable to analyze and intuitive to monitor. The concept of platoons will be proposed in Section \ref{sec:platooning}; in this section, we focus on air highways.

Let an air highway be denoted by the continuous function $\hw: [0, 1] \rightarrow \mathbb{R}^2$. Such a highway lies in a horizontal plane of fixed altitude, with start and end points given by $\hw(0)\in\mathbb{R}^2$ and $\hw(1)\in\mathbb{R}^2$ respectively. For simplicity, we assume that the highway segment is a straight line segment, and the parameter $s$ indicates the position in some fixed altitude as follows: $\hw(s) = \hw(0) + s(\hw(1) - \hw(0))$. To each highway, we assign a speed of travel $v_\hw$ and specify the direction of travel to be the direction from $\hw(0)$ to $\hw(1)$, denoted using a unit vector $\hwd = \frac{\hw(1) - \hw(0)}{\lVert\hw(1) - \hw(0)\rVert_2}$. As we will show in Section \ref{sec:platooning}, UAVs use simple controllers to track the highway.

Air highways must not only provide structure to make the analysis of a large number of vehicles tractable, but also allow vehicles to reach their destinations while minimizing any relevant costs to the vehicles and to the surrounding regions. Such costs can for example take into account people, assets on the ground, and manned aviation, entities to which UAVs pose the biggest risks \cite{Kopardekar16}. Thus, given an origin-destination pair (eg. two cities), air highways must connect the two points while potentially satisfying other criteria. In addition, optimal air highway locations should ideally be able to be recomputed in real-time when necessary in order to update airspace constraints on-the-fly, in case, for example, airport configurations change or certain airspaces have to be closed \cite{Kopardekar16}. With this in mind, we now define the air highway placement problem, and propose a simple and fast way to approximate its solution that allows for real-time recomputation. Our solution based on solving the Eikonal equation can be thought of as converting a cost map over a geographic area in continuous space into a discrete graph whose nodes are waypoints joined by edges which are the air highways.

Note that the primary purpose of this section is to provide a method for the real-time placement of air highways. The specifics of determining the cost map based on population density, geography, weather forecast information, etc., as well as the criteria for when air highway locations need to be updated, is beyond the scope of this paper.

In addition, if vehicles in the airspace are far away from each other, it may be reasonable for all vehicles to fly in an unstructured manner. As long as multiple-way conflicts do not occur, pairwise collision avoidance maneuvers would be sufficient to ensure safety. Unstructured flight is likely to result in more efficient trajectories for each individual vehicle. However, whether multiple-way conflicts occur cannot be predicted ahead of time, and are not guaranteed to be resolvable when they occur. By organizing vehicles into platoons, the likelihood of multiple-way conflicts is vastly reduced. Structured flight is in general less efficient for the individual vehicle, and this loss of efficiency can be thought of as a cost incurred by the vehicles in order ensure higher levels of safety.

In general, there may be many different levels of abstractions in the airspace. For larger regions such as cities, air highways may prove beneficial, and for a small region such as a neighborhood, perhaps unstructured flight is sufficiently safe. Further research is needed to better understand parameters such as the density of vehicles above which unstructured flight is no longer manageable, and other details like platoon size.

\subsection{The Air Highway Placement Problem}
Consider a map $\cmap:\mathbb{R}^2 \rightarrow \mathbb{R}$ which defines the cost $\cmap(\pos)$ incurred when a UAV flies over the position $\pos=(\pos_x,\pos_y)\in\mathbb{R}^2$. Given any position $\pos$, a large value of $\cmap(\pos)$ indicates that the position $\pos$ is costly or undesirable for a UAV to fly over. Locations with high cost could, for example, include densely populated areas and areas around airports. In general, the cost map $\cmap(\cdot)$ may be used to model cost of interference with commercial airspaces, cost of accidents, cost of noise pollution, risks incurred, etc., and can be flexibly specified by government regulation bodies. 

Let $\pos^o$ denote an origin point and $\pos^d$ denote a destination point. Consider a sequence of highways $\hws_N = \{\hw_1, \hw_2, \ldots, \hw_N\}$ that satisfies the following:

\begin{equation}
\label{eq:hw_seq}
\begin{aligned}
\hw_1(0) &= \pos^o \\
\hw_i(1) &= \hw_{i+1}(0), i = 0, 1, \ldots, N-1 \\
\hw_N(1) &= \pos^d \\
\end{aligned}
\end{equation}

The interpretation of the above conditions is that the start point of first highway is the origin, the end point of a highway is the start point of the next highway, and the end point of last highway is the destination. The highways $\hw_1,\ldots,\hw_N$ form a sequence of waypoints for a UAV starting at the origin $\pos^o$ to reach its destination $\pos^d$.

Given only the origin point $\pos^o$ and destination point $\pos^d$, there are an infinite number of choices for a sequence of highways that satisfy \eqref{eq:hw_seq}. However, if one takes into account the cost of flying over any position $\pos$ using the cost map $\cmap(\cdot)$, we arrive at the air highway placement problem:

\begin{equation}
\label{eq:ahpp} 
\begin{aligned}
& \min_{\hws_N, N} \left\{\left(\sum_{i=1}^N \int_0^1 \cmap(\hw_i(s)) ds\right) + R(N)\right\}\\
& \text{subject to \eqref{eq:hw_seq}} 
\end{aligned}
\end{equation}

\noindent where $R(\cdot)$ is a regularizer, such as $R(N) = N^2$.

The interpretation of \eqref{eq:ahpp} is that we consider air highways to be line segments of constant altitude over a region, and UAV platoons travel on these air highways to get from some origin to some destination. Any UAV flying on a highway over some position $\pos$ incurs a cost of $\cmap(\pos)$, so that the total cost of flying from the origin to the destination is given by the summation in \eqref{eq:ahpp}. The air highway placement problem minimizes the cumulative cost of flying from some origin $p^o$ to some destination $p^d$ along the sequence of highways $\hws_N=\{\hw_1, \hw_2, \ldots, \hw_N\}$. The regularization term $R(N)$ is used to prevent $N$ from being arbitrarily large.

\begin{figure}[h!]
	\centering
	\includegraphics[width=0.95\columnwidth]{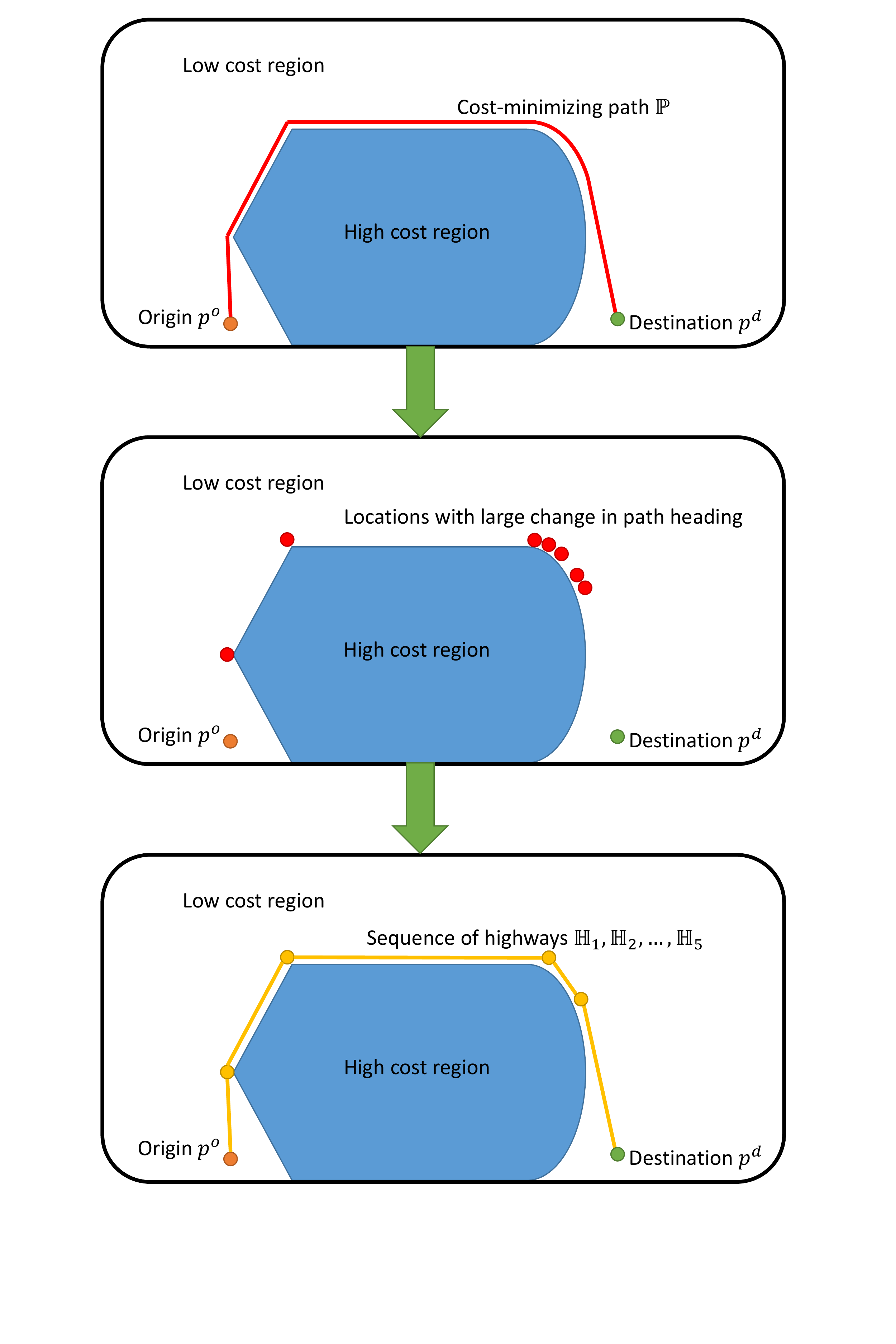}
	\caption{Illustration of the air highway placement procedure.}
	\label{fig:hw_ill}
\end{figure}
\subsection{The Eikonal Equation -- Cost-Minimizing Path}
Let $s_0, s_1\in \mathbb{R}$, and let $\ppath: [s_0, s_1] \rightarrow \mathbb{R}^2$ be a path starting from an origin point $\pos^o = \ppath(s_0)$ and ending at a destination point $\pos^d = \ppath(s_1)$. Note that the sequence $\hws_N$ in \eqref{eq:ahpp} is a piecewise affine example of a path $\ppath(s), s\in[s_0, s_1]$; however, a path $\ppath$ that is not piecewise affine cannot be written as a sequence of highways $\hws_N$.

More concretely, suppose a UAV flies from an origin point $p^o$ to a destination point $p^d$ along some path $\ppath(s)$ parametrized by $s$. Then, $\ppath(s_0) = p^o$ would denote the origin, and $\ppath(s_1) = p^d$ would denote the destination. All intermediate $s$ values denote the intermediate positions of the path, i.e. $\ppath(s) = \pos(s) = (\pos_x(s), \pos_y(s))$.

Consider the cost map $\cmap(\pos_x, \pos_y)$ which captures the cost incurred for UAVs flying over the position $\pos = (\pos_x, \pos_y)$. Along the entire path $\ppath(s)$, the cumulative cost $\ccost(\ppath)$ is incurred. Define $\ccost$ as follows:

\begin{equation}
\ccost(\ppath) = \int_{s_0}^{s_1} \cmap(\ppath(s)) ds
\end{equation}

For an origin-destination pair, we would like to find the path such that the above cost is minimized. More generally, given an origin point $p^o$, we would like to compute the function $\ocost$ representing the optimal cumulative cost for any destination point $\pos^d$:

\begin{equation}
\label{eq:rahpp} 
\begin{aligned}
\ocost(\pos^d) &= \min_{\ppath(\cdot), \ppath(s_1)=\pos^d} \ccost(\ppath) \\
&= \min_{\ppath(\cdot), \ppath(s_1)=\pos^d} \int_{s_0}^{s_1} \cmap(\ppath(s)) ds
\end{aligned}
\end{equation}

It is well known that the viscosity solution \cite{Crandall84} to the Eikonal equation \eqref{eq:eikonal} precisely computes the function $\ocost(\pos^d)$ given the cost map $\cmap$ \cite{Sethian96,Alton06}. Note that a single function characterizes the minimum cost from an origin $\pos^o$ to \textit{any} destination $\pos^d$. Once $\ocost$ is found, the optimal path $\ppath$ between $\pos^o$ and $\pos^d$ can be obtained via gradient descent.

\begin{equation}
\label{eq:eikonal}
\begin{aligned}
\cmap(\pos)|\nabla \ocost(\pos)| &= 1 \\
\ocost(\pos^o) &= 0
\end{aligned}
\end{equation}

The Eikonal equation \eqref{eq:eikonal} can be efficiently computed numerically using the fast marching method \cite{Sethian96}; each computation takes on the order of merely a second.

Note that \eqref{eq:rahpp} can be viewed as a relaxation of the air highway placement problem defined in \eqref{eq:ahpp}. Unlike \eqref{eq:ahpp}, the relaxation \eqref{eq:rahpp} can be quickly solved using currently available numerical tools. Thus, we first solve the approximate air highway placement problem \eqref{eq:rahpp} by solving \eqref{eq:eikonal}, and then post-process the solution to \eqref{eq:rahpp} to obtain an approximation to \eqref{eq:ahpp}.

Given a single origin point $\pos^o$, the optimal cumulative cost function $\ocost(\pos^d)$ can be computed. Suppose $M$ different destination points $\pos^d_i,i=1,\ldots,M$ are chosen. Then, $M$ different optimal paths $\ppath_i,i=1,\ldots,M$ are obtained from $\ocost$. 
\subsection{From Paths to Waypoints}
Each of the cost-minimizing paths $\ppath_i$ computed from the solution to the Eikonal equation consists of a closely-spaced set of points. Each path $\ppath_i$ is an approximation to the sequence of highways $\hws_{N_i}^i = \{\hw^i_j\}_{i=1,j=1}^{i=M,j=N_i}$ defined in \eqref{eq:ahpp}, but now indexed by the corresponding path index $i$. 

For each path $\ppath_i$, we would like to sparsify the points on the path to obtain a collection of waypoints, $\wpt_{i,j}, j = 1,\ldots, N_i+1$, which are the end points of the highways:

\begin{equation}
\begin{aligned}
\hw^i_j(0) &= \wpt_{i,j}, \\
\hw^i_j(1) &= \wpt_{i,j+1}, \\
j &= 1,\ldots,N_i
\end{aligned}
\end{equation}

There are many different ways to do this, and this process will not be our focus. However, for illustrative purposes, we show how this process may be started. We begin by noting the path's heading at the destination point. We add to the collection of waypoints the first point on the path at which the heading changes by some threshold $\theta_C$, and repeat this process along the entire path.

If there is a large change in heading within a small section of the cost-minimizing path, then the collection of points may contain many points which are close together. In addition, there may be multiple paths that are very close to each other (in fact, this behavior is desirable), which may contribute to cluttering the airspace with too many waypoints. To reduce clutter, one could cluster the points. Afterwards, each cluster of points can be replaced by a single point located at the centroid of the cluster. 

To the collection of points resulting from the above process, we add the origin and destination points. Repeating the entire process for every path, we obtain waypoints for all the cost-minimizing paths under consideration. Figure \ref{fig:hw_ill} summarizes the entire air highway placement process, including our example of how the closely-spaced set of points on a path can be sparsified.
\subsection{Results}
To illustrate our air highway placement proposal, we used the San Francisco Bay Area as an example, and classified each point on the map into four different regions: ``regions around airports'', ``highly populated cities'', ``water'', and ``other''. Each region has an associated cost, reflecting the desirability of flying a vehicle over an area in the region. In general, these costs can be arbitrary and determined by government regulation agencies. For illustration purposes, we assumed the following categories and costs:

\begin{itemize}
\item Region around airports: $\cost{airports}=b$,
\item Cities: $\cost{cities}=1$,
\item Water: $\cost{water}=b^{-2}$,
\item Other: $\cost{other}=b^{-1}$.
\end{itemize}

This assumption assigns costs in descending order to the categories ``regions around airports'', ``cities'', ``other'', and ``water''. Flying a UAV in each category is more costly by a factor of $b$ compared to the next most important category. The factor $b>1$ is a tuning parameter that we adjusted to vary the relative importance of the different categories, and we used $b=4$ in the figures below.

Figure \ref{fig:airHighway_results} shows the San Francisco Bay Area geographic map, cost map, cost-minimizing paths, and contours of the value function $V$. The region enclosed by the black boundary represents ``region around airports'', which have the highest cost. The dark blue, yellow, and light blue regions represent the ``cities'', the ``water'', and the ``other''' categories, respectively. We assumed that the origin corresponds to the city ``Concord'', and chose a number of other major cities as destinations.

A couple of important observations can be made here. First, the cost-minimizing paths to the various destinations in general overlap, and only split up when they are very close to entering their destination cities. This is intuitively desirable because having overlapping cost-minimizing paths keeps the number of distinct air highways low. Secondly, the contours, which correspond to level curves of the value function, have a spacings corresponding to the cost map: the spacings are large in areas of low cost, and vice versa. This provides insight into the placement of air highways to destinations that were not shown in this example.

\begin{figure*}[h!]
	\centering
	\includegraphics[width=\textwidth]{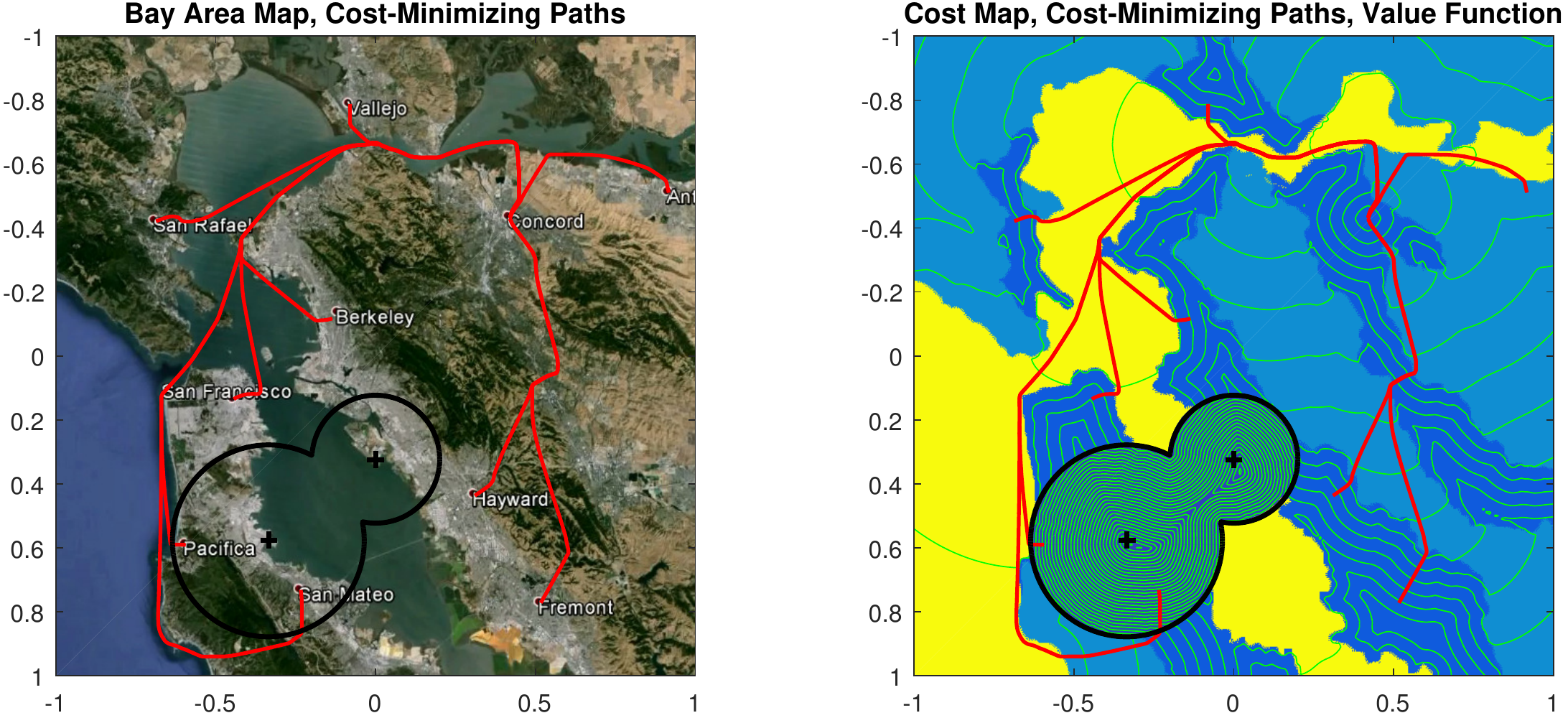}
	\caption{Cost-minimizing paths computed by the Fast Marching Method based on the assumed cost map of the San Francisco Bay Area.}
	\label{fig:airHighway_results}
\end{figure*}

\begin{figure*}[h!]
	\centering
	\includegraphics[width=\textwidth]{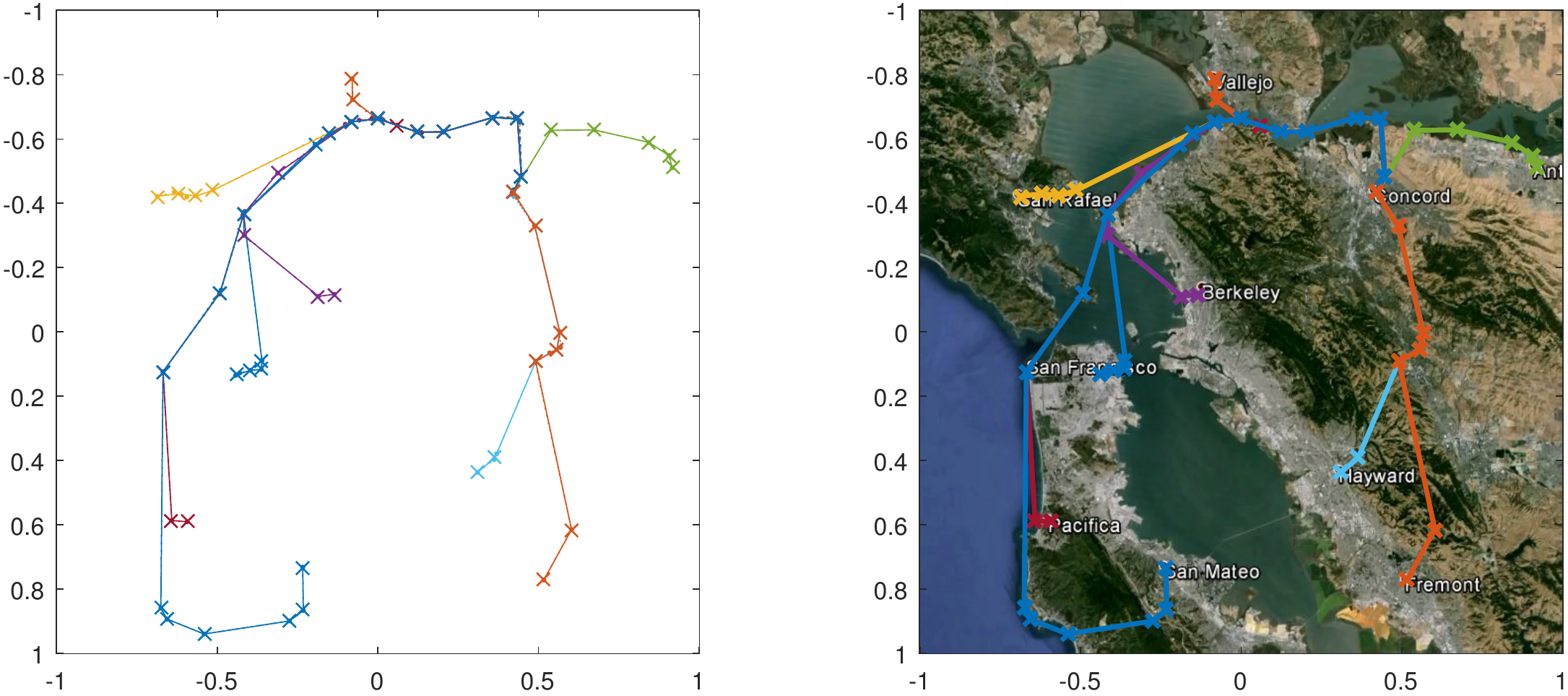}
	\caption{Results of conversion from cost-minimizing paths to highway way points.}
	\label{fig:airHighway_sparse}
\end{figure*}

Figure \ref{fig:airHighway_sparse} shows the result of converting the cost-minimizing paths to a small number of waypoints. The left plot shows the waypoints, interpreted as the start and end points of air highways, over a white background for clarity. The right plot shows these air highways over the map of the Bay Area. Note that we could have gone further to merge some of the overlapping highways. However, the purpose of this section is to illustrate the natural occurrence of air highways from cost-minimizing paths; post-processing of the cost-minimizing paths, which serve as a guide for defining air highways, is not our focus. 

\subsection{Real-Time Highway Location Updates}
Since \eqref{eq:eikonal} can be solved in approximately 1 second, the air highway placement process can be redone in real-time if the cost map changes at a particular time. This flexibility can be useful in any situation in which unforeseen circumstances could cause a change in the cost of a particular region of the airspace. For example, accidents or disaster response vehicles may result in an area temporarily having a high cost. On the other hand, depending on for instance the time of day, it may be most desirable to fly in different regions of the airspace, resulting in those regions temporarily having a low cost.
\section{Unmanned Aerial Vehicle Platooning \label{sec:platooning}}
Air highways exhibiting trunk routes that separate near destinations motivate the use of platoons which fly on these highways. The air highway structure along with the UAV platooning concept together enable the use of reachability to analyze safety and goal satisfaction properties. The structure reduces the likelihood of multiple-way conflicts, and makes pairwise analysis more indicative of the joint safety of all UAVs. In addition to reducing complexity, the proposed structure is intuitive, and allows human participation in the monitoring and management of the unmanned airspace.

Organizing UAVs into platoons implies that the UAVs cannot fly in an unstructured way, and must have a restricted set of controllers or maneuvers depending on the UAV's role in the airspace. To model UAVs flying in platoons on air highways, we propose a hybrid system whose modes of operations describe a UAV's role in the highway structure. For the hybrid system model, reachability analysis is used to enable successful and safe operation and mode transitions.

\subsection{UAVs in Platoons}
\subsubsection{Vehicle Dynamics}
Consider a UAV whose dynamics are given by
\begin{equation}
\label{eq:veh_dyn}
\dot{x} = f(x,u)
\end{equation}

\noindent where $x$ represents the state, and $u$ represents the control action. The techniques we present in this paper do not depend on the dynamics of the vehicles, as long as their dynamics are known. However, for concreteness, we assume that the UAVs are quadrotors that fly at a constant altitude under non-faulty circumstances. For the quadrotor, we use a simple model in which the $x$ and $y$ dynamics are double integrators:

\begin{equation}
\label{eq:dyn}
\begin{aligned}
\dot{\pos}_x &= \vel_x \\
\dot{\pos}_y &= \vel_y  \\
\dot{\vel}_x &= u_x \\
\dot{\vel}_y &= u_y \\
|u_x|,|u_y| &\le u_\text{max}
\end{aligned}
\end{equation}

\noindent where the state $x=(\pos_x, \vel_x, \pos_y, \vel_y)\in\mathbb{R}^4$ represents the quadrotor's position in the $x$-direction, its velocity in the $x$-direction, and its position and velocity in the $y$-direction, respectively. The control input $u = (u_x, u_y)\in\mathbb{R}^2$ consists of the acceleration in the $x$- and $y$- directions. For convenience, we will denote the position and velocity $\pos=(\pos_x, \pos_y),\vel=(\vel_x,\vel_y)$, respectively. 

In general, the problem of collision avoidance among $N$ vehicles cannot be tractably solved using traditional dynamic programming approaches because the computation complexity of these approaches scales exponentially with the number of vehicles. Thus, in our present work, we will consider the situation where UAVs travel on air highways in platoons, defined in the following sections. The structure imposed by air highways and platooning enables us to analyze the safety and goal satisfaction properties of the vehicles in a tractable manner.

\subsubsection{Vehicles as Hybrid Systems}
We model each vehicle as a hybrid system \cite{Lygeros98, Lygeros12} consisting of the modes ``Free", ``Leader", ``Follower", and ``Faulty". Within each mode, a vehicle has a set of restricted maneuvers, including one that allows the vehicle to change modes if desired. The modes and maneuvers are as follows:

\begin{itemize}
\item Free: 

A Free vehicle is not in a platoon or on a highway, and its possible maneuvers or mode transitions are
\begin{itemize}
\item remain a Free vehicle by staying away from highways
\item become a Leader by entering a highway to create a new platoon
\item become a Follower by joining a platoon that is currently on a highway
\end{itemize} 

\item Leader: 

A Leader vehicle is the vehicle at the front of a platoon (which could consist of only the vehicle itself). The available maneuvers and mode transitions are

\begin{itemize}
\item remain a Leader by traveling along the highway at a pre-specified speed $\vel_\hw$
\item become a Follower by merging the current platoon with a platoon in front
\item become a Free vehicle by leaving the highway
\end{itemize}

\item Follower: 

A Follower vehicle is a vehicle that is following a platoon leader. The available maneuvers and mode transitions are 

\begin{itemize}
\item remain a Follower by staying a distance of $d_\text{sep}$ behind the vehicle in front in the current platoon
\item remain a Follower by joining a different platoon on another highway
\item become a Leader by splitting from the current platoon while remaining on the highway
\item become a Free vehicle by leaving the highway
\end{itemize}

\item Faulty: 

If a vehicle from any of the other modes becomes unable to operate within the allowed set of maneuvers, it transitions into the Faulty mode. Reasons for transitioning to the Faulty mode include vehicle malfunctions, performing collision avoidance with respect to another Faulty vehicle, etc. A Faulty vehicle is assumed to descend via a fail-safe mechanism after some pre-specified duration $\td$ to a different altitude level where it no longer poses a threat to vehicles on the air highway system.

Such a fail-safe mechanism could be an emergency landing procedure such as those analyzed in \cite{Adler2012,Coombes2014,Idicula2015}. Typically, emergency landing involves identifying the type of fault and finding feasible landing locations given the dynamics during the fault. We will omit these details and summarize them into $\td$, the time required to exit the current altitude level.
\end{itemize}

\begin{figure}
	\centering
	\includegraphics[width=\columnwidth]{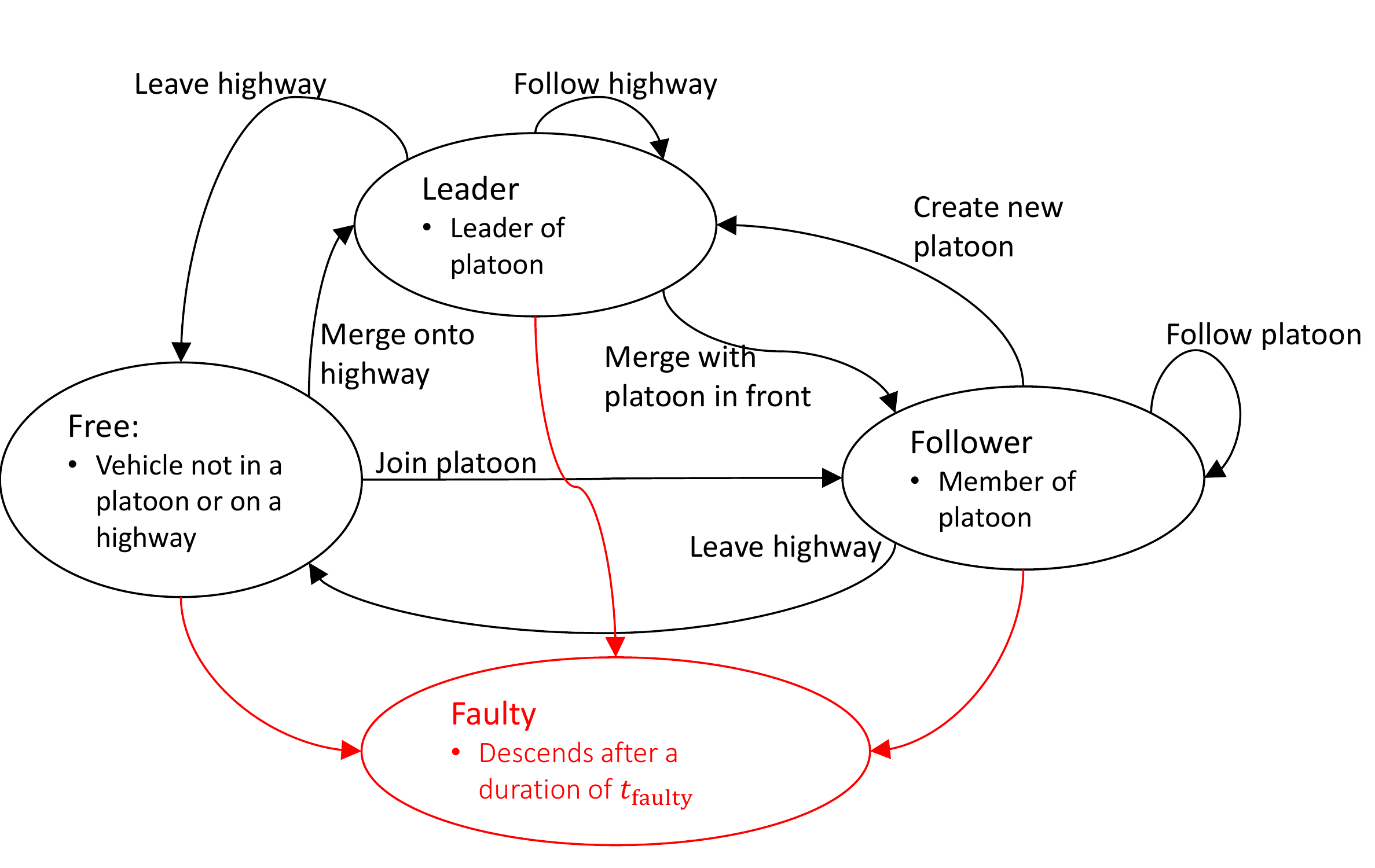}
	\caption{Hybrid modes for vehicles in platoons. Vehicles begin in the ``Free'' mode before they enter the highway.}
	\label{fig:vehicleModes}
\end{figure}

The available maneuvers and associated mode transitions are summarized in Figure \ref{fig:vehicleModes}.

Suppose that there are $N$ vehicles in total in the airspace containing the highway system. We will denote the $N$ vehicles as $\veh{i}, i=1\ldots,N$. We consider a platoon of vehicles to be a group of $M$ vehicles ($M\le N$), denoted $\veh{P_1}, \ldots, \veh{P_M}, \{P_j\}_{j=1}^M \subseteq \{i\}_{i=1}^N$, in a single-file formation. When necessary, we will use superscripts to denote vehicles of different platoons: $\veh{P_i^j}$ represents the $i$th vehicle in the $j$th platoon. 

\begin{figure}
	\centering
	\includegraphics[width=\columnwidth]{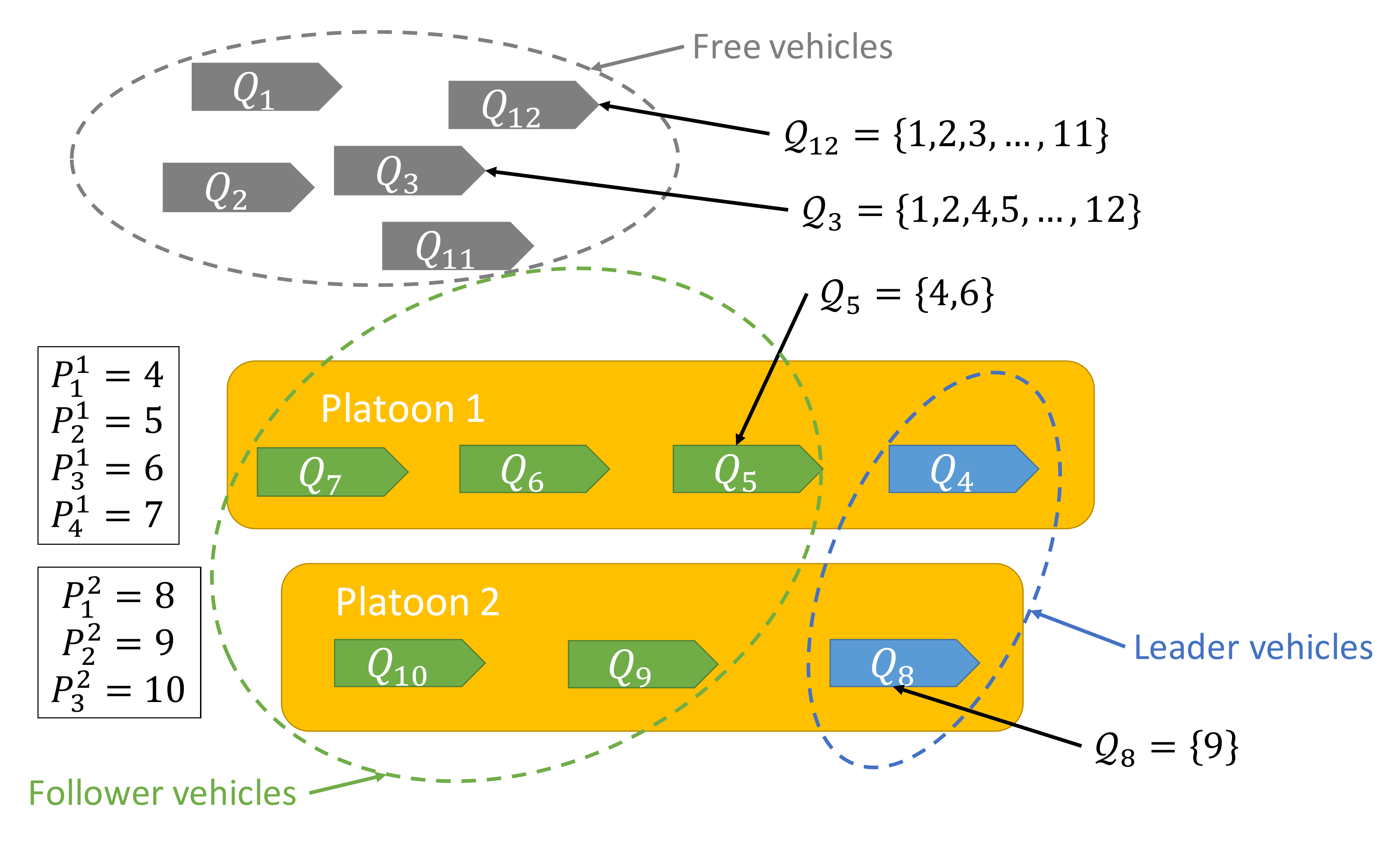}
	\caption{Notation for vehicles in platoons.}
	\label{fig:vehicleNotation}
\end{figure}

For convenience, let $\vehSCS{i}$ denote the set of indices of vehicles with respect to which $Q_i$ checks safety. If vehicle $\veh{i}$ is a free vehicle, then it must check for safety with respect to all other vehicles, $\vehSCS{i} = \{j: j\neq i\}$. If the vehicle is part of a platoon, then it checks safety with respect to the platoon member in front and behind, $\vehSCS{i} = \{P_{j+1}, P_{j-1}\}$. Figure \ref{fig:vehicleNotation} summarizes the indexing system of the vehicles.

We will organize the vehicles into platoons travel along air highways. The vehicles maintain a separation distance of $\sepdist$ with their neighbors inside the platoon. In order to allow for close proximity of the vehicles and the ability to resolve multiple simultaneous safety breaches, we assume that when a vehicle exhibits unpredictable behavior, it will be able to exit the altitude range of the highway within a duration of $\td$. Such a requirement may be implemented practically as an fail-safe mechanism to which the vehicles revert when needed.

\subsubsection{Objectives}
Given the above modeling assumptions, our goal is to provide control strategies to guarantee the success and safety of all the mode transitions. The theoretical tool used to provide the safety and goal satisfaction guarantees is reachability. The BRSs we compute will allow each vehicle to perform complex actions such as 

\begin{itemize}
\item merge onto a highway to form a platoon
\item join a new platoon
\item leave a platoon to create a new one
\item react to malfunctioning or intruder vehicles
\end{itemize}

We also propose more basic controllers to perform other simpler actions such as
\begin{itemize}
\item follow the highway at constant altitude at a specified speed
\item maintain a constant relative position and velocity with respect to the leader of a platoon
\end{itemize}

In general, the control strategy of each vehicle has a safety component, which specifies a set of states that it must avoid, and a goal satisfaction component, which specifies a set of states that the vehicle aims to reach. Together, the safety and goal satisfaction controllers guarantee the safety and success of a vehicle in the airspace making any desired mode transition. In this paper, these guarantees are provided using reachability analysis, and allow the multi-UAV system to perform joint maneuvers essential to maintaining structure in the airspace.
\subsection{Hamilton-Jacobi Reachability}
\subsubsection{General Framework}
Consider a differential game between two players described by the system
\begin{equation}
\dot{x} = f(x, u_1, u_2), \text{for almost every }t\in [-T,0]
\end{equation}

\noindent where $x\in\mathbb{R}^n$ is the system state, $u_1\in \mathcal{U}_1$ is the control of Player 1, and $u_2\in\mathcal{U}_2$ is the control of Player 2. We assume $f:\mathbb{R}^n\times \mathcal{U}_1 \times \mathcal{U}_2 \rightarrow \mathbb{R}^n$ is uniformly continuous, bounded, and Lipschitz continuous in $x$ for fixed $u_1,u_2$, and the control functions $u_1(\cdot)\in\mathbb{U}_1,u_2(\cdot)\in\mathbb{U}_2$ are drawn from the set of measurable functions\footnote{
A function $f:X\to Y$ between two measurable spaces $(X,\Sigma_X)$ and $(Y,\Sigma_Y)$ is said to be measurable if the preimage of a measurable set in $Y$ is a measurable set in $X$, that is: $\forall V\in\Sigma_Y, f^{-1}(V)\in\Sigma_X$, with $\Sigma_X,\Sigma_Y$ $\sigma$-algebras on $X$,$Y$.}. Player 2 is allowed to use nonanticipative strategies \cite{Evans84,Varaiya67} $\gamma$, defined by

\begin{equation}
\begin{aligned}
\gamma &\in \Gamma := \{\mathcal{N}: \mathbb{U}_1 \rightarrow \mathbb{U}_2 \mid  u_1(r) = \hat{u}_1(r) \\
&\text{for almost every } r\in[t,s] \Rightarrow \mathcal{N}[u_1](r) \\
&= \mathcal{N}[\hat{u}_1](r) \text{ for almost every } r\in[t,s]\}
\end{aligned}
\end{equation}

In our differential game, the goal of Player 2 is to drive the system into some target set $\mathcal{L}$, and the goal of Player 1 is to drive the system away from it. The set $\mathcal{L}$ is represented as the zero sublevel set of a bounded, Lipschitz continuous function $l:\mathbb{R}^n\rightarrow\mathbb{R}$. We call $l(\cdot)$ the \textit{implicit surface function} representing the set $\mathcal L: \mathcal{L}=\{x\in\mathbb{R}^n \mid l(x)\le 0\}$.

Given the dynamics \eqref{eq:dyn} and the target set $\mathcal{L}$, we would like to compute the BRS, $\mathcal{V}(t)$:

\begin{equation}
\begin{aligned}
\mathcal{V}(t) &:= \{x\in\mathbb{R}^n \mid \exists \gamma\in\Gamma \text{ such that } \forall u_1(\cdot)\in\mathbb{U}_1, \\
&\exists s \in [t,0], \xi_f(s; t, x, u_1(\cdot), \gamma[u_1](\cdot)) \in \mathcal{L} \}
\end{aligned}
\end{equation}
where $\xi_f$ is the trajectory of the system satisfying initial conditions $\xi_f(t; x, t, u_1(\cdot), u_2(\cdot))=x$ and the following differential equation almost everywhere on $[-t, 0]$
\begin{equation}
\begin{aligned}
\frac{d}{ds}&\xi_f(s; x, t, u_1(\cdot), u_2(\cdot)) \\
&= f(\xi_f(s; x, t, u_1(\cdot), u_2(\cdot)), u_1(s), u_2(s))
\end{aligned}
\end{equation}

Many methods involving solving HJ PDEs \cite{Mitchell05} and HJ variational inequalities (VI) \cite{Bokanowski10,Barron89,Fisac15} have been developed for computing the BRS. These HJ PDEs and HJ VIs can be solved using well-established numerical methods. For this paper, we use the formulation in \cite{Mitchell05}, which shows that the BRS $\mathcal{V}(t)$ can be obtained as the zero sublevel set of the viscosity solution \cite{Crandall84} $V(t,x)$ of the following terminal value HJ PDE:

\begin{equation} \label{eq:HJIPDE}
\begin{aligned}
D_t &V(t,x) + \\
&\min \{0, \max_{u_1\in\mathcal{U}_1} \min_{u_2\in\mathcal{U}_2} D_x V(t,x) \cdot f(x,u_1,u_2) \} = 0, \\
&V(0,x) = l(x)
\end{aligned}
\end{equation}

\begin{figure}
	\centering
	\includegraphics[width=\columnwidth]{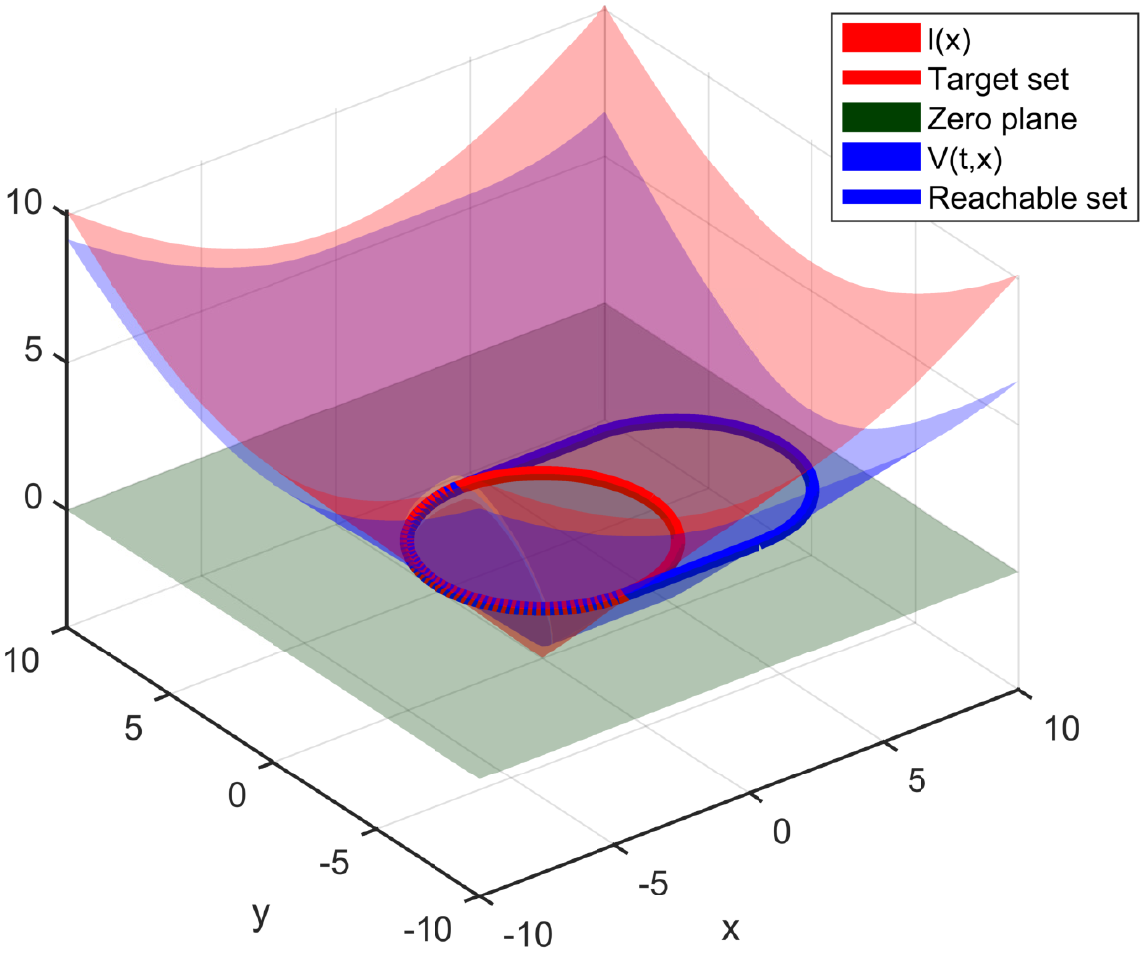}
	\caption{Illustration of a target set, a BRS, and their implicit surface functions.}
	\label{fig:RSExample}
\end{figure}

\noindent from which we obtain $\mathcal{V}(t) = \{x\in\mathbb{R}^n \mid V(t,x)\le 0\}$. The target set, reachable set, and implicit surface functions $l(x)$ and $V(t,x)$ representing them are shown in Fig. \ref{fig:RSExample}. From the solution $V(t,x)$, we can also obtain the optimal controls for both players via the following:

\begin{equation} \label{eq:HJI_ctrl_syn}
\begin{aligned}
u_1^*(t,x) &= \arg \max_{u_1\in\mathcal{U}_1} \min_{u_2\in\mathcal U_2} D_x V(t,x) \cdot f(x,u_1,u_2)\\
u_2^*(t,x) &= \arg \min_{u_2\in\mathcal{U}_2} D_x V(t,x) \cdot f(x,u_1^*,u_2)
\end{aligned}
\end{equation}

In the special case where there is only one player, we obtain an optimal control problem for a system with dynamics

\begin{equation} \label{eq:dyn_d}
\dot{x} = f(x, u), t\in [-T,0], u\in\mathcal U.
\end{equation}

The BRS in this case would be given by the HJ PDE

\begin{equation} \label{eq:HJBPDE}
\begin{aligned}
D_t V(t,x) + \min \{0, \min_{u\in\mathcal{U}} D_x V(t,x) \cdot f(x,u)\} &= 0 \\
V(0,x) = l(x)&
\end{aligned}
\end{equation}

\noindent where the optimal control is given by

\begin{equation} \label{eq:HJB_ctrl_syn}
u^*(t,x) = \arg \min_{u\in\mathcal{U}} D_x V(t,x) \cdot f(x,u)
\end{equation}

For our application, we will use several decoupled system models and utilize decomposition techniques \cite{Chen15, Chen17b, Chen2016a}, which enables real-time 4D BRS computations and tractable 6D BRS computations.

\subsubsection{Relative Dynamics and Augmented Relative Dynamics}
Besides Equation \eqref{eq:dyn}, we will also consider the relative dynamics between two quadrotors $Q_i,Q_j$. These dynamics can be obtained by defining the relative variables

\begin{equation} \label{eq:rel_var}
\begin{aligned}
p_{x,r} &= p_{x,i} - p_{x,j} \\
p_{y,r} &= p_{y,i} - p_{y,j}\\
v_{x,r} &= v_{x,i} - v_{x,j}\\
v_{y,r} &= v_{y,i} - v_{y,j}
\end{aligned}
\end{equation}

We treat $Q_i$ as Player 1, the evader who wishes to avoid collision, and we treat $Q_j$ as Player 2, the pursuer, or disturbance, that wishes to cause a collision. In terms of the relative variables given in \eqref{eq:rel_var}, we have 

\begin{equation} \label{eq:rel_dyn}
\begin{aligned}
\dot{p}_{x,r} &= v_{x,r} \\
\dot{p}_{y,r} &= v_{y,r} \\
\dot{v}_{x,r} &= u_{x,i} - u_{x,j} \\
\dot{v}_{y,r} &= u_{y,i} - u_{y,j}
\end{aligned}
\end{equation}


We also augment \eqref{eq:rel_var} with the velocity of $Q_i$, given in \eqref{eq:rel_dyn_aug}, to impose a velocity limit on the quadrotor.

\begin{equation} \label{eq:rel_dyn_aug}
\begin{aligned}
\dot{p}_{x,r} &= v_{x,r} \\
\dot{p}_{y,r} &= v_{y,r} \\
\dot{v}_{x,r} &= u_{x,i} - u_{x,j} \\
\dot{v}_{y,r} &= u_{y,i} - u_{y,j} \\
\dot{v}_{x,i} &= u_{x,i} \\
\dot{v}_{y,i} &= u_{y,i}
\end{aligned}
\end{equation}
\subsection{Reachability-Based Controllers \label{sec:reach_ctrl}}
Reachability analysis is useful for constructing controllers in a large variety of situations. In order to construct different controllers, an appropriate target set needs to be defined depending on the goal of the controller. If one defines the target set to be a set of desired states, the BRS would represent the states that a system needs to first arrive at in order to reach the desired states. On the other hand, if the target set represents a set of undesirable states, then the BRS would indicate the region of the state space that the system needs to avoid. In addition, the system dynamics with which the BRS is computed provide additional flexibility when using reachability to construct controllers.

Using a number of different target sets and dynamics, we now propose different reachability-based controllers used for vehicle mode transitions in our platooning concept.

\subsubsection{Getting to a Target State \label{sec:abs_target_ctrl}}
The controller used by a vehicle to reach a target state is important in two situations in the platooning context. First, a vehicle in the ``Free'' mode can use the controller to merge onto a highway, forming a platoon and changing modes to a ``Leader'' vehicle. Second, a vehicle in either the ``Leader'' mode or the ``Follower'' mode can use this controller to change to a different highway, becoming a ``Leader'' vehicle. 

In both of the above cases, we use the dynamics of a single vehicle specified in \eqref{eq:dyn}. The target state would be a position $(\Pos_x, \Pos_y)$ representing the desired merging point on the highway, along with a velocity $(\Vel_x, \Vel_y)=\vel_\hw$ that corresponds to the speed and direction of travel specified by the highway. For the reachability computation, we define the target set to be a small range of states around the target state $\bar x_H = (\Pos_x, \Pos_y, \Vel_x, \Vel_y)$:

\begin{equation}
\begin{aligned}
\mathcal{L}_H = \{x: |\pos_x-\Pos_x|\le r_{\pos_x}, |v_x-\Vel_x|\le r_{\vel_x}, \\
|\pos_y - \Pos_y| \le r_{\pos_y}, |v_y - \Vel_y|\le r_{\vel_y} \}.
\end{aligned}
\end{equation}

Here, we represent the target set $\mathcal{L}_H$ as the zero sublevel set of the function $l_H(x)$, which specifies the terminal condition of the HJ PDE that we need to solve. Once the HJ PDE is solved, we obtain the BRS $\mathcal V_H(t)$ from the subzero level set of the solution $V_H(t,x)$. More concretely, $\mathcal{V}_H(T) = \{x: V_H(-T,x)\le 0\}$ is the set of states from which the system can be driven to the target $\mathcal{L}_H$ within a duration of $T$. 

Depending on the time horizon $T$, the size of the BRS $\mathcal V_H(T)$ varies. In general, a vehicle may not initially be inside the BRS $\mathcal V_H(T)$, yet it needs to be in order to get to its desired target state. Determining a control strategy to reach $\mathcal V_H(T)$ is itself a reachability problem (with $\mathcal V_H(T)$ as the target set), and it would seem like this reachability problem needs to be solved in order for us to use the results from our first reachability problem. However, practically, one could choose $T$ to be large enough to cover a sufficiently large area to include any practically conceivable initial state. From our simulations, a suitable algorithm for getting to a desired target state is as follows:

\begin{enumerate}
\item Move towards $\bar{x}_H$ in pure pursuit with some velocity, until $V_H(-T,x)\le 0$. In practice, this step consistently drives the system into the BRS.
\item Apply the optimal control extracted from $V_H(-T,x)$ according to \eqref{eq:HJB_ctrl_syn} until $\mathcal{L}_H$ is reached.
\end{enumerate}

\subsubsection{Getting to a State Relative to Another Vehicle \label{sec:rel_target_ctrl}}
In the platooning context, being able to go to a state relative to another moving vehicle is important for the purpose of forming and joining platoons. For example, a ``Free'' vehicle may join an existing platoon that is on a highway and change modes to become a ``Follower''. Also, a ``Leader'' or ``Follower'' may join another platoon and afterwards go into the ``Follower'' mode.

To construct a controller for getting to a state relative to another vehicle, we use the relative dynamics of two vehicles, given in \eqref{eq:rel_dyn}. In general, the target state is specified to be some position $(\Pos_{x,r}, \Pos_{y,r})$ and velocity $(\Vel_{x,r}, \Vel_{y,r})$ relative to a reference vehicle. In the case of a vehicle joining a platoon that maintains a single file, the reference vehicle would be the platoon leader, the desired relative position would be a certain distance behind the leader, depending on how many other vehicles are already in the platoon; the desired relative velocity would be $(0,0)$ so that the formation can be kept.

For the reachability problem, we define the target set to be a small range of states around the target state $\bar x_P = (\Pos_{x,r}, \Pos_{y,r}, \Vel_{x,r}, \Vel_{y,r})$:

\begin{equation}
\begin{aligned}
\mathcal{L}_P = \{x: |\pos_{x,r}-\Pos_{x,r}|\le r_{\pos_x}, |\vel_{x,r}-\Vel_{x,r}|\le r_{\vel_x}, \\
|\pos_{y,r} - \Pos_{y,r}| \le r_{\pos_y}, |\vel_{y,r} - \Vel_{y,r}|\le r_{\vel_y} \}
\end{aligned}
\end{equation}

The target set $\mathcal{L}_P$ is represented by the zero sublevel set of the implicit surface function $l_P(x)$, which specifies the terminal condition of the HJ PDE \eqref{eq:HJIPDE}. The zero sublevel set of the solution to \eqref{eq:HJIPDE}, $V_P(-T,x)$, gives us the set of relative states from which a quadrotor can reach the target in the relative coordinates within a duration of $T$. In the BRS computation, we assume that the reference vehicle moves along the highway at constant speed, so that $u_j(t)$ = 0. The following is a suitable algorithm for a vehicle joining a platoon to follow the platoon leader:

\begin{enumerate}
\item Move towards $\bar{x}_P$ in a straight line, with some velocity, until $V_P(-T,x)\le 0$.
\item Apply the optimal control extracted from $V_P(-T,x)$ according to \eqref{eq:HJI_ctrl_syn} until $\mathcal{L}_P$ is reached.
\end{enumerate}

\subsubsection{Avoiding Collisions \label{sec:collision_ctrl}}
A vehicle can use a goal satisfaction controller described in the previous sections when it is not in any danger of collision with other vehicles. If the vehicle could potentially be involved in a collision within the next short period of time, it must switch to a safety controller. The safety controller is available in every mode, and executing the safety controller to perform an avoidance maneuver does not change a vehicle's mode. 

In the context of our platooning concept, we define an unsafe configuration as follows: a vehicle is either within a minimum separation distance $d$ to a reference vehicle in both the $x$ and $y$ directions, or is traveling with a speed above the speed limit $\vel_\text{max}$ in either of the $x$ and $y$ directions. To take this specification into account, we use the augmented relative dynamics given by \eqref{eq:rel_dyn_aug} for the reachability problem, and define the target set as follows:

\begin{equation}
\begin{aligned}
\mathcal{L}_S = \{x: &|\pos_{x,r}|, |\pos_{y,r}|\le d \vee |\vel_{x,i}| \ge \vel_\text{max} \vee |\vel_{y,i}| \ge \vel_\text{max} \}
\end{aligned}
\end{equation}

We can now define the implicit surface function $l_S(x)$ corresponding to $\mathcal{L}_S$, and solve the HJ PDE \eqref{eq:HJIPDE} using $l_S(x)$ as the terminal condition. As before, the zero sublevel set of the solution $V_S(t,x)$ specifies the BRS $\mathcal{V}_S(t)$, which characterizes the states in the augmented relative coordinates, as defined in \eqref{eq:rel_dyn_aug}, from which $Q_i$ \textit{cannot} avoid $\mathcal{L}_S$ for a time period of $t$, if $Q_j$ uses the worst-case control. To avoid collisions, $Q_i$ must apply the safety controller according to \eqref{eq:HJI_ctrl_syn} on the boundary of the BRS in order to avoid going into the BRS. The following algorithm wraps our safety controller around goal satisfaction controllers:

\begin{enumerate}
\item For a specified time horizon $t$, evaluate $V_S(-t,x_i-x_j)$ for all $j\in \mathcal{Q}(i)$.

$\mathcal{Q}(i)$ is the set of quadrotors with which vehicle $\veh{i}$ checks safety.
\item Use the safety or goal satisfaction controller depending on the values $V_S(-t,x_i-x_j),j\in \mathcal{Q}(i)$: 

If $\exists j\in \mathcal{Q}(i),V_S(-t,x_i-x_j)\le 0$, then $Q_i,Q_j$ are in potential conflict, and $Q_i$ must use a safety controller; otherwise $Q_i$ may use a goal satisfaction controller.
\end{enumerate}

\subsection{Other Controllers \label{sec:other_ctrl}}
Reachability was used in Section \ref{sec:reach_ctrl} for the relatively complex maneuvers that require safety and goal satisfaction guarantees. For the simpler maneuvers of traveling along a highway and following a platoon, many well-known classical controllers suffice. For illustration, we use the simple controllers described below.

\subsubsection{Traveling along a highway} \label{sec:travel_hwy}
We use a model-predictive controller (MPC) for traveling along a highway; this controller allows the leader to travel along a highway at a pre-specified speed. Here, the goal is for a leader vehicle to track an air highway $\hw(s), s\in[0,1]$ while maintaining some constant velocity $\vel_\hw$ specified by the highway. The highway and the specified velocity can be written as a desired position and velocity over time, $\bar \pos(t), \bar \vel (t)$. Assuming that the initial position on the highway, $s_0=s(t_0)$ is specified, such a controller can be obtained from the following optimization problem over the time horizon $[t_0, t_1]$:

\begin{equation}
\begin{aligned}
\text{minimize } & \int_{t_0}^{t_1} \big\{\| p(t)-\bar\pos(t) \|_2 + \\ 
&\qquad \| v(t) - \bar\vel(t) \|_2 + 1-s \big\} dt \\
\text{subject to } & \text{vehicle dynamics } \eqref{eq:veh_dyn} \\
& |u_x|, |u_y| \le u_\text{max}, |v_x|, |v_y| \le v_\text{max} \\
& s(t_0) = s_0, \dot{s} \ge 0
\end{aligned}
\end{equation}

If we discretize time, the above optimization is becomes convex optimization over a small number of decision variables, and can be quickly solved.

\subsubsection{Following a Platoon} \label{sec:follow_platoon}
Follower vehicles use a feedback control law tracking a nominal position and velocity in the platoon, with an additional feedforward term given by the leader's acceleration input; here, for simplicity, we assume perfect communication between the leader and the follower vehicles. This following law enables smooth vehicle trajectories in the relative platoon frame, while allowing the platoon as a whole to perform agile maneuvers by transmitting the leader's acceleration command $u_{P_1}(t)$ to all vehicles.

The $i$-th member of the platoon, $Q_{P_i}$, is expected to track a relative position in the platoon $r^i = (r_x^i,r_y^i)$ with respect to the leader's position $p_{P_1}$, and the leader's velocity $v_{P_1}$ at all times. The resulting control law has the form:
\begin{equation}\label{eq:follow}
u^i(t) = k_p \big[p_{P_1}(t) + r^i(t) - p^i(t) \big] + k_v\big[v_{P_1}(t) - v^i(t)\big] + u_{P_1}(t)
\end{equation}
for some $k_p,k_v>0$. In particular, a simple rule for determining $r^i(t)$ in a single-file platoon is given for $Q_{P_i}$ as:
\begin{equation}\label{eq:nominal_pos}
r^i(t) = - (i-1) \sepdist \hwd
\end{equation}
where $\sepdist$ is the spacing between vehicles along the platoon and $\hwd$ is the highway's direction of travel.
\subsection{Summary of Controllers}
\begin{figure}
	\centering
	\includegraphics[width=0.75\columnwidth]{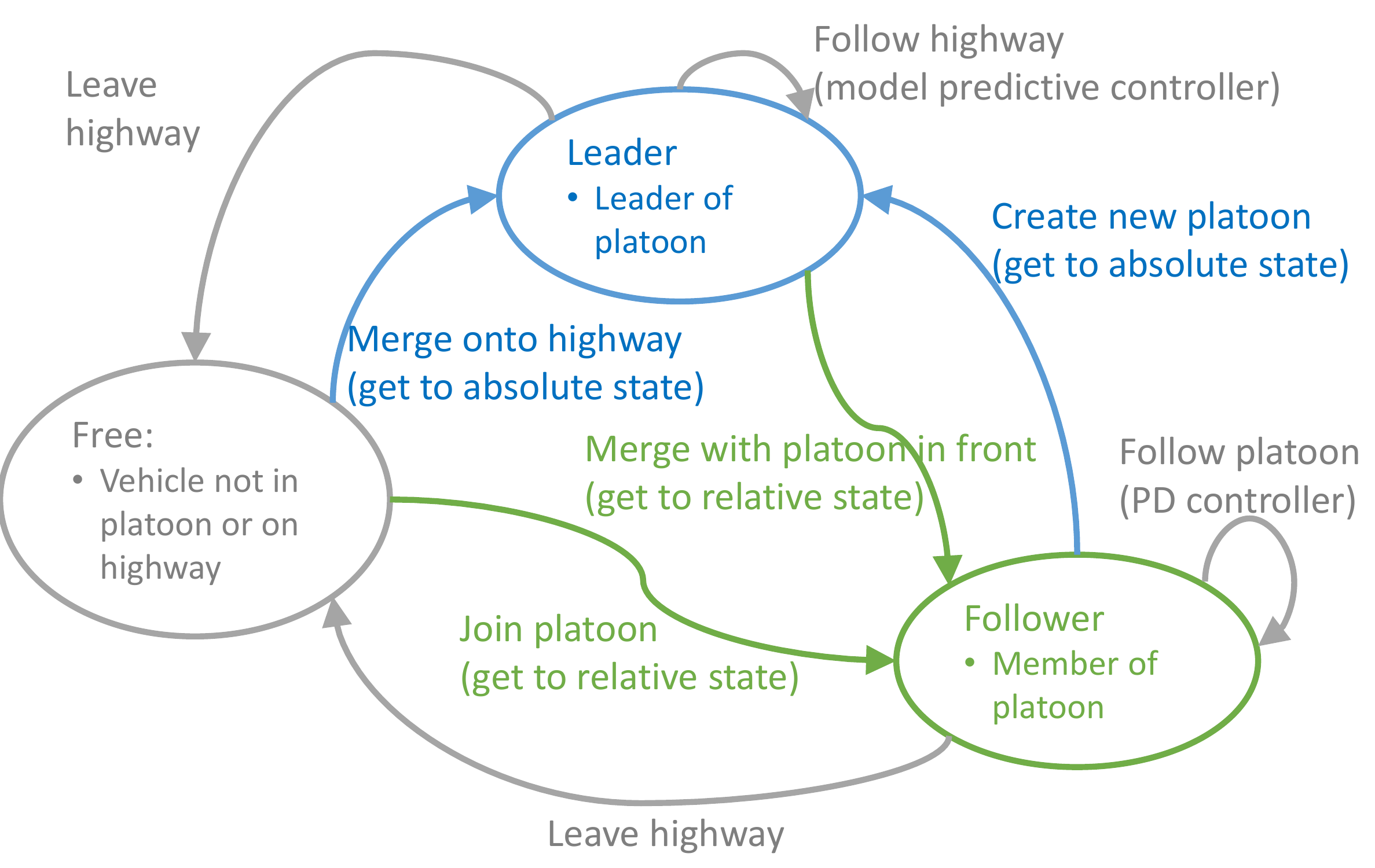}
	\caption{Summary of mode switching controllers. Reachability-based controllers are shown as the blue and green arrows.}
	\label{fig:modeControllers}
\end{figure}

We have introduced several reachability-based controllers, as well as some simple controllers. Pairwise collision avoidance is guaranteed using the safety controller, described in Section \ref{sec:platooning}-\ref{sec:reach_ctrl}-\ref{sec:collision_ctrl}. As long as a vehicle is not in potential danger according to the safety BRSs, it is free to use any other controller. All of these other controllers are \textit{goal satisfaction controllers}, and their corresponding mode transitions are shown in Figure \ref{fig:modeControllers}.

The controller for getting to an absolute target state, described in Section \ref{sec:platooning}-\ref{sec:reach_ctrl}-\ref{sec:abs_target_ctrl}, is used whenever a vehicle needs to move onto a highway to become a platoon leader. This controller guarantees the success of the mode transitions shown in blue in Figure \ref{fig:modeControllers}.

The controller for getting to a relative target state, described in Section \ref{sec:platooning}-\ref{sec:reach_ctrl}-\ref{sec:rel_target_ctrl}, is used whenever a vehicle needs to join a platoon to become a follower. This controller guarantees the success of the mode transitions shown in green in Figure \ref{fig:modeControllers}.

For the simple maneuvers of traveling along a highway or following a platoon, many simple controllers such as the ones suggested in Section \ref{sec:platooning}-\ref{sec:other_ctrl} can be used. These controllers keep the vehicles in either the Leader or the Follower mode. Alternatively, additional controllers can be designed for exiting the highway, although these are not considered in this paper. All of these non-reachability-based controllers are shown in gray in Figure \ref{fig:modeControllers}.

\subsection{Safety Analysis}
Under normal operations in a single platoon, each follower vehicle $\veh{i},i=P_2,\ldots,P_{M-1}$ in a platoon checks whether it is in the safety BRS with respect to $\veh{P_{i-1}}$ and $\veh{P_{i+1}}$. So $\vehSCS{i} = \{P_{i+1}, P_{i-1}\}$ for $i=P_2,\ldots,P_{N-1}$. Assuming there are no nearby vehicles outside of the platoon, the platoon leader $\veh{P_1}$ checks safety against $\veh{P_2}$, and the platoon trailer $Q_{P_N}$ checks safety against $Q_{P_{N-1}}$. So $\vehSCS{P_1}=\{P_2\}, \vehSCS{P_N}=\{P_{N-1}\}$. When all vehicles are using goal satisfaction controllers to perform their allowed maneuvers, in most situations no pair of vehicles should be in an unsafe configuration. However, occasionally a vehicle $\veh{k}$ may behave unexpectedly due to faults or malfunctions, in which case it may come into an unsafe configuration with another vehicle.

With our choice of $\vehSCS{i}$ and the assumption that the platoon is in a single-file formation, some vehicle $\veh{i}$ would get near the safety BRS with respect to $\veh{k}$, where $\veh{k}$ is likely to be the vehicle in front or behind of $\veh{i}$. In this case, a ``safety breach" occurs. Our synthesis of the safety controller guarantees the following: between every pair of vehicles $\veh{i},\veh{k}$, if $V_S(-t,x_i- x_k)>0$, then $\exists u_i$ to keep $\veh{i}$ from colliding with $\veh{k}$ for a desired time horizon $t$, despite the worst case (an adversarial) control from $\veh{k}$. Therefore, as long as the number of ``safety breaches" is at most one for $\veh{i}$, $Q_i$ can simply use the optimal control to avoid $\veh{k}$ and avoid collision for the time horizon of $t$. Under the assumption that vehicles are able to exit the current altitude range within a duration of $\td$, if we choose $t=\td$, the safety breach would always end before any collision can occur. 

Within a duration of $\td$, there is a small chance that additional safety breaches may occur. However, as long as the total number of safety breaches does not exceed the number of affected vehicles, collision avoidance of all the vehicles can be guaranteed for the duration $\td$. However, as our simulation results show, placing vehicles in single-file platoons makes the likelihood of multiple safety breaches low during the presence of one intruder vehicle. 

In the event that multiple safety breaches occur for some of the vehicles due to a malfunctioning vehicle within the platoon or intruding vehicles outside of the platoon, vehicles that are causing safety breaches must exit the highway altitude range in order to avoid collisions. Every extra altitude range reduces the number of simultaneous safety breaches by $1$, so $K$ simultaneous safety breaches can be resolved using $K-1$ different altitude ranges. The general process and details of the complete picture of multi-altitude collision avoidance is part of our future work. 

The concept of platooning can be coupled with any collision avoidance algorithm that provides safety guarantees. In this paper, we have only proposed the simplest reachability-based collision avoidance scheme. Existing collision avoidance algorithms such as \cite{Bansal16} and \cite{Chen16} have the potential to provide safety guarantees for many vehicles in order to resolve multiple safety breaches at once. Coupling the platooning concept with the more advanced collision avoidance methods that provide guarantees for a larger number of vehicles would reduce the risk of multiple safety breaches.

Given that vehicles within a platoon are safe with respect to each other, each platoon can be treated as a single vehicle, and perform collision avoidance with other platoons when needed. The option of treating each platoon as a single unit can reduce the number of individual vehicles that need to check for safety against each other, reducing overall computation burden.
\subsection{Numerical Simulations}
In this section, we consider several situations that vehicles in a platoon on an air highway may commonly encounter, and show via simulations the behaviors that emerge from the controllers we defined in Sections \ref{sec:platooning}-\ref{sec:reach_ctrl} and \ref{sec:platooning}-\ref{sec:other_ctrl}.

\subsubsection{Forming a Platoon}
We first consider the scenario in which Free vehicles merge onto an initially unoccupied highway. In order to do this, each vehicle first checks safety with respect to all other vehicles, and uses the safety controller if necessary, according to Section \ref{sec:platooning}-\ref{sec:reach_ctrl}-\ref{sec:collision_ctrl}. Otherwise, the vehicle uses the goal satisfaction controller for getting to an absolute target set described in Section \ref{sec:platooning}-\ref{sec:reach_ctrl}-\ref{sec:abs_target_ctrl} in order to merge onto the highway, create a platoon, and become a Leader vehicle if there are no platoons on the highway. If there is already a platoon on the highway, then the vehicle would use the goal satisfaction controller for getting to a target set relative to the platoon leader as described in Section \ref{sec:platooning}-\ref{sec:reach_ctrl}-\ref{sec:rel_target_ctrl} to join the platoon and become a Follower.

\begin{figure*}[h!]
    \centering
    \begin{subfigure}[t]{0.9\columnwidth} \label{subfig:fp_43}
        \includegraphics[width=\columnwidth]{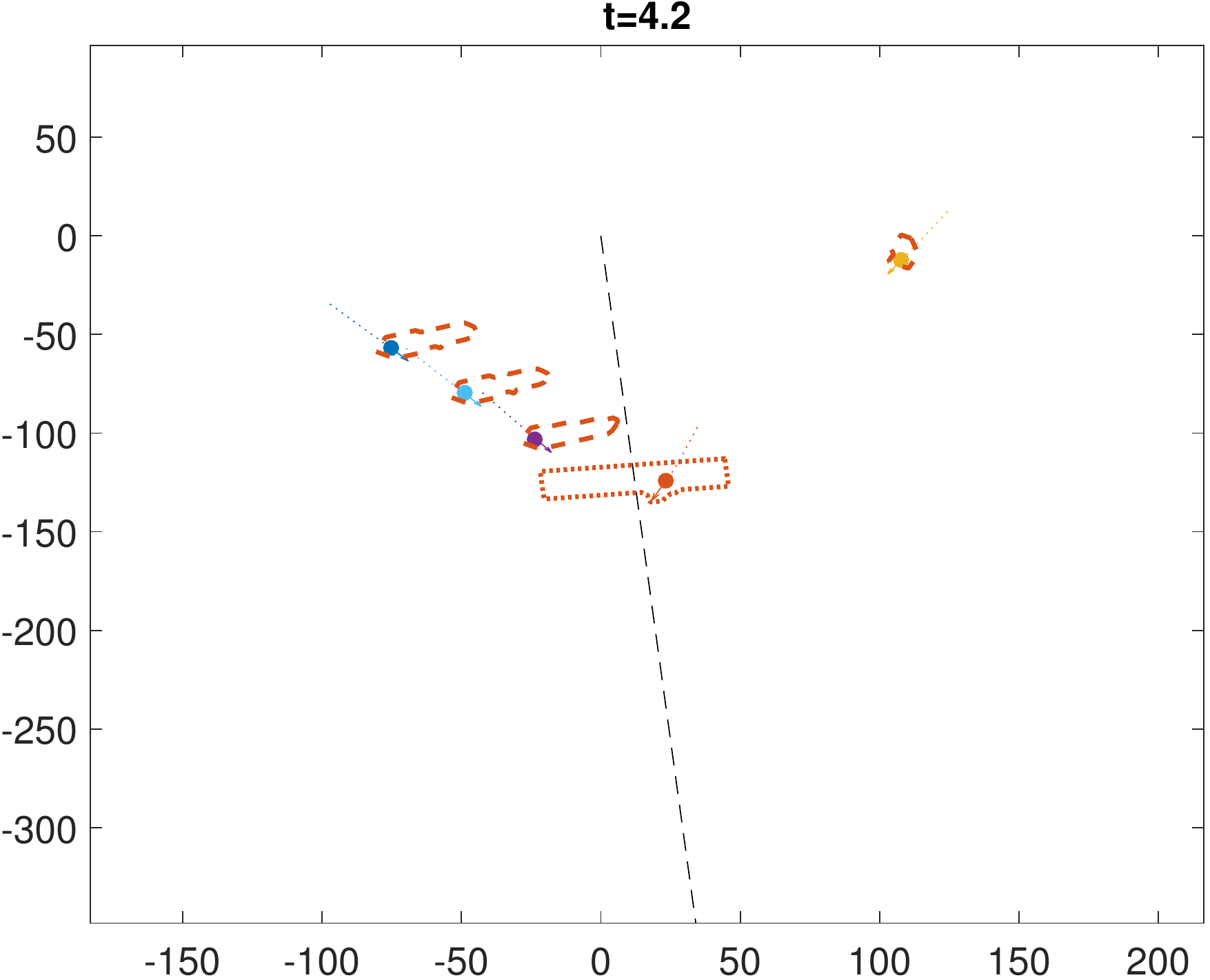}
        \caption{The red vehicle is merging onto the highway while avoiding collisions.}
    \end{subfigure}
    \begin{subfigure}[t]{0.9\columnwidth} \label{subfig:fp_81}
        \includegraphics[width=\columnwidth]{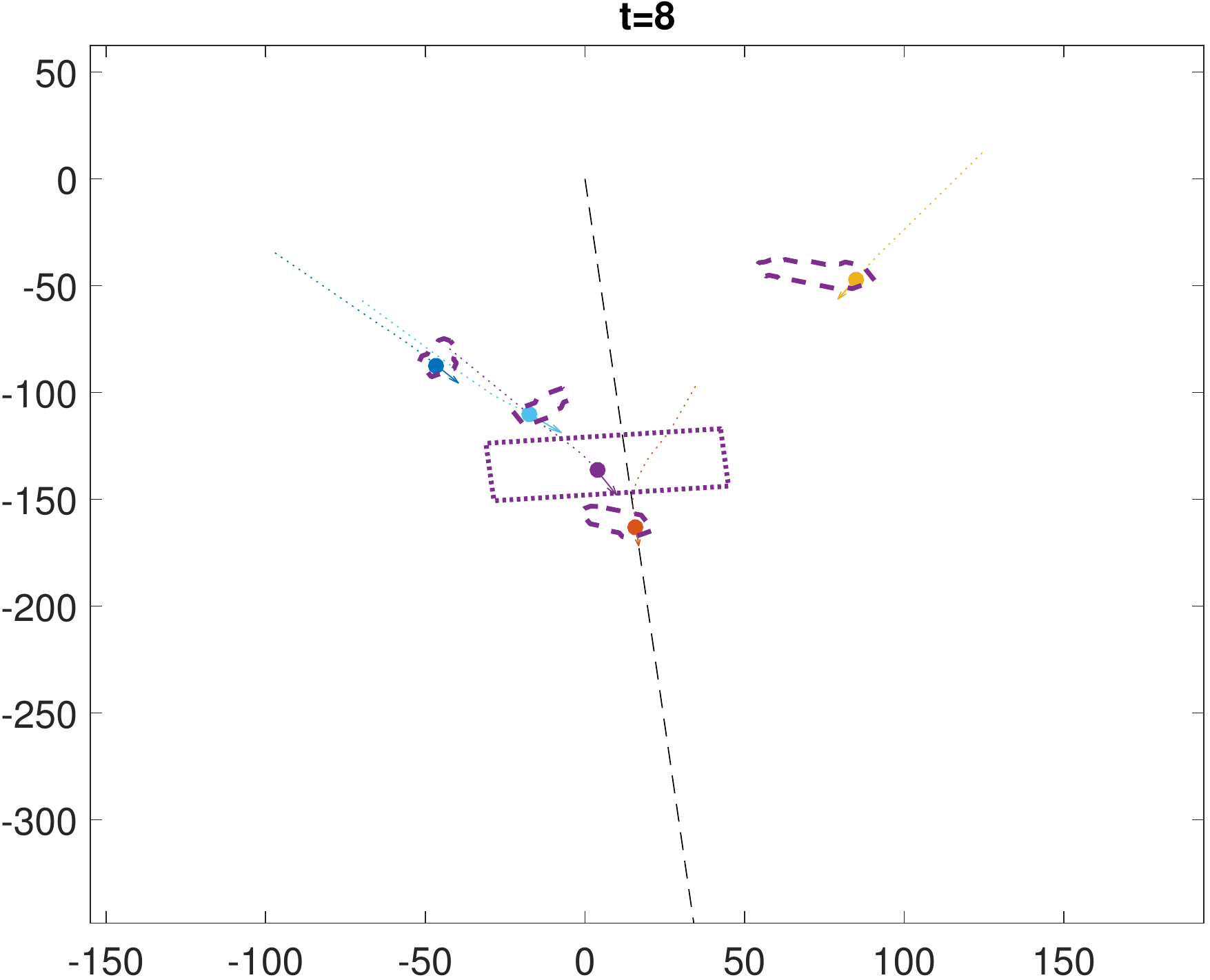}
        \caption{The purple vehicle is joining the platoon while avoiding collisions.}
    \end{subfigure}

    \begin{subfigure}[t]{0.9\columnwidth} \label{subfig:fp_110}
        \includegraphics[width=\columnwidth]{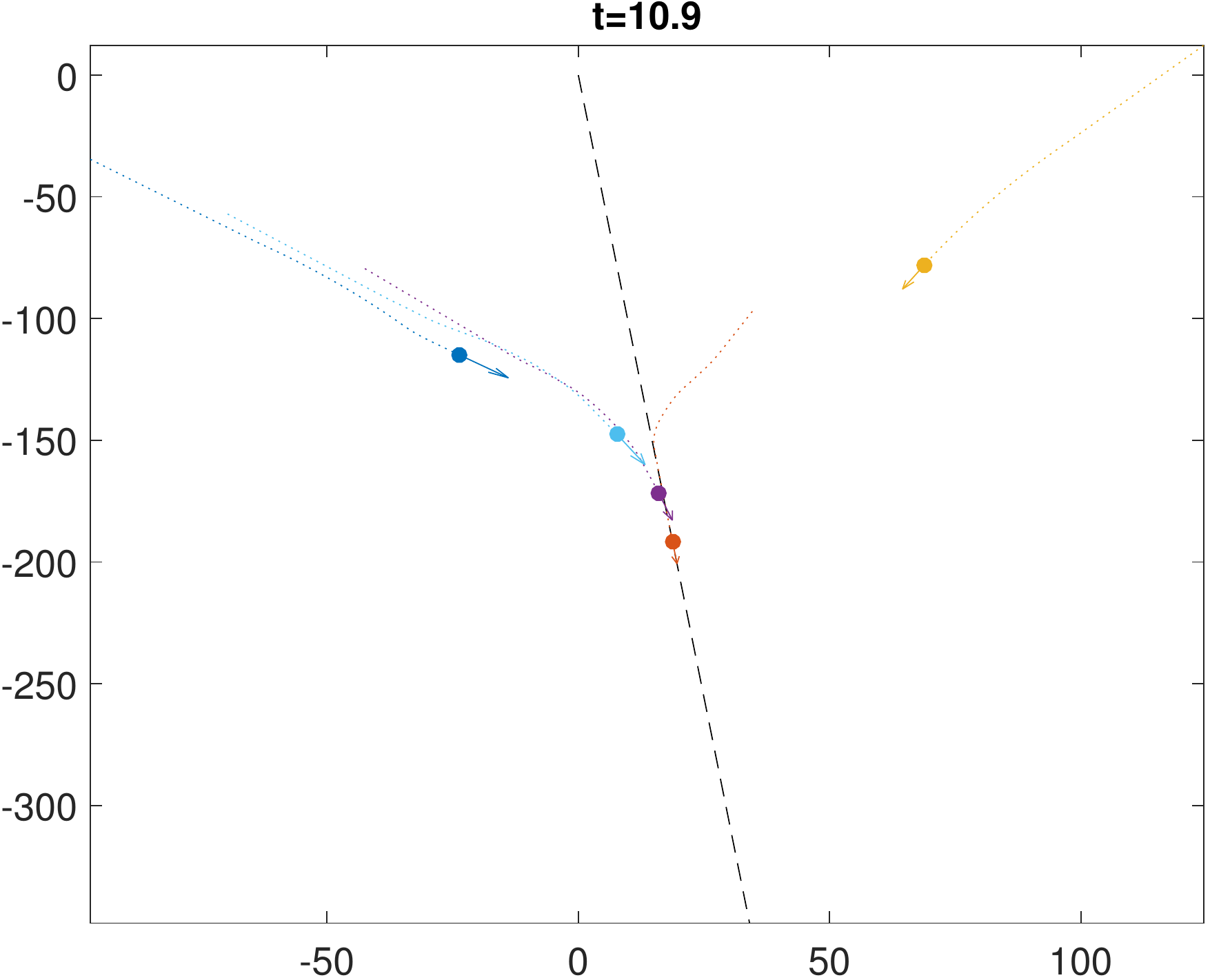}
        \caption{The last three vehicles follow the same process to join the platoon.}
    \end{subfigure}
    \begin{subfigure}[t]{0.9\columnwidth} \label{subfig:fp_210}
        \includegraphics[width=\columnwidth]{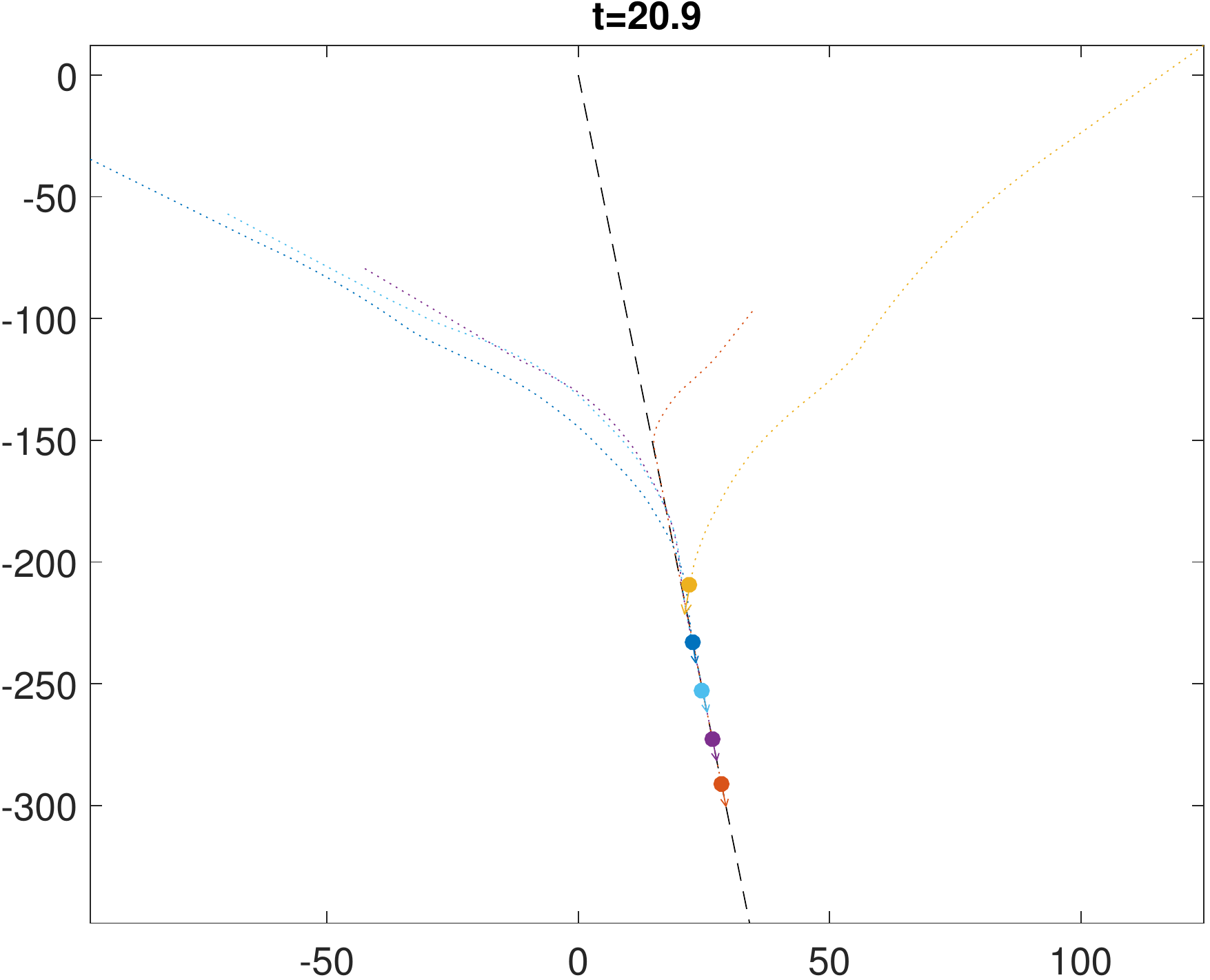}
        \caption{All five vehicles have successfully joined the platoon and now travel on the highway together.}
    \end{subfigure}   
    \caption{A simulation showing how five vehicles initially in the Free mode can form a platoon. \label{fig:fp}}
\end{figure*}

For the simulation example, shown in Figure \ref{fig:fp}, the highway is specified by a line segment beginning at the origin. The five vehicles, $\veh{1}, \veh{2}, \ldots, \veh{5}$ are colored orange, purple, light blue, dark blue, and yellow, respectively.

The first two plots in Figure \ref{fig:fp} illustrate the use of safety and goal satisfaction BRS for the first two vehicles. Since the goal satisfaction BRSs are in 4D and the safety BRSs are in 6D, we compute and plot their 2D slices based on the vehicles' velocities and relative velocities.  All vehicles begin as Free vehicles, so they each need to take into account five different BRSs: four safety BRSs and one goal satisfaction BRS. For clarity, we only show the goal satisfaction BRS and the four safety BRSs for one of the vehicles. 

For $\veh{1}$ (orange), an arbitrary point of entry on the highway is chosen as the target absolute position, and the velocity corresponding to a speed of $10$ m/s in the direction of the highway is chosen as the target absolute velocity. This forms the target state $\bar{x}_H=(\bar{p}_x, \bar{v}_x, \bar{p}_y, \bar{v}_y)$, from which we define the target set $\mathcal{L}_H$ as in Section \ref{sec:platooning}-\ref{sec:reach_ctrl}-\ref{sec:abs_target_ctrl}.

At $t=4.2$, $\veh{1}$ (orange) is inside the goal satisfaction BRS for getting to an absolute state, shown as the dotted orange boundary. Therefore, it is ``locked-in" to the target state $\bar{x}_H$, and follows the optimal control in \eqref{eq:HJB_ctrl_syn} to $\bar{x}_H$. During the entire time, $\veh{1}$ checks whether it may collide with any of the other vehicles within a time horizon of $\td$. To do this, it simply checks whether its state relative to each of the other vehicles is within the corresponding safety BRS. As an example, the safety BRS boundary with respect to $\veh{2}$ (purple) is shown as the orange dashed boundary around $\veh{2}$ (purple); $\veh{1}$ (orange) is safe with respect to $\veh{2}$ (purple) since $\veh{1}$ (orange) is outside of the boundary. In fact, $\veh{1}$ is safe with respect to all vehicles.

After completing merging onto the empty highway, $\veh{1}$ (orange) creates a platoon and becomes its leader, while subsequent vehicles begin to form a platoon behind the leader in the order of ascending distance to $\veh{1}$ (orange) according to the process described in Section \ref{sec:platooning}-\ref{sec:reach_ctrl}-\ref{sec:rel_target_ctrl}. Here, we choose the target relative position $(\bar{p}_{x,r}, \bar{p}_{y,r})$ to be a distance $\sepdist$ behind the last reserved slot in the platoon, and the target relative velocity $(\bar{v}_{x,r}, \bar{v}_{y,r}) = (0,0)$ with respect to the leader in order to maintain the platoon formation. This gives us the target set $\mathcal{L}_P$ that we need.

At $t=8.0$, $Q_2$ (purple) is in the process of joining the platoon behind $\veh{1}$ (orange) by moving towards the target $\bar{x}_P$ relative to the position of $\veh{1}$ (orange). Note that $\bar{x}_P$ moves with $\veh{1}$ (orange) as $\bar{x}_P$ is defined in terms of the relative states of the two vehicles. Since $\veh{2}$ is inside the goal satisfaction BRS boundary for joining the platoon (purple dotted boundary), it is ``locked-in" to the target relative state $\bar{x}_P$, and begins following the optimal control in \eqref{eq:HJI_ctrl_syn} towards the target as long as it stays out of all safety BRSs. For example, at $t=5.9$, $\veh{2}$ (purple) is outside of the safety BRS with respect to $\veh{1}$ (orange), shown as the purple dashed boundary around $\veh{1}$ (orange). Again, from the other safety BRS boundaries, we can see that $\veh{2}$ is in fact safe with respect to all vehicles.

In the bottom plots of Figure \ref{fig:fp}, $\veh{1}$ (orange) and $\veh{2}$ (purple) have already become the platoon leader and follower, respectively. The rest of the vehicles follow the same process to join the platoon. All 5 vehicles eventually form a single platoon and travel along the highway together. As with the first two vehicles, the goal satisfaction controllers allow the remaining vehicles to optimally and smoothly join the platoon, while the safety controllers prevent collisions from occurring.

\subsubsection{Intruder Vehicle}
We now consider the scenario in which a platoon of vehicles encounters an intruder vehicle. To avoid collision, each vehicle checks for safety with respect to the intruder and any vehicles in front and behind of it in the platoon. If necessary, the vehicle uses the reachability-based safety controller to avoid collision, otherwise it uses the appropriate controller to travel on the highway if it is a leader, or follow the leader if it is a follower. After danger has passed, the vehicles in the platoon resume normal operation.

\begin{figure*}[h!]
    \centering
    \begin{subfigure}[t]{0.9\columnwidth} \label{subfig:in_100}
        \includegraphics[width=\columnwidth]{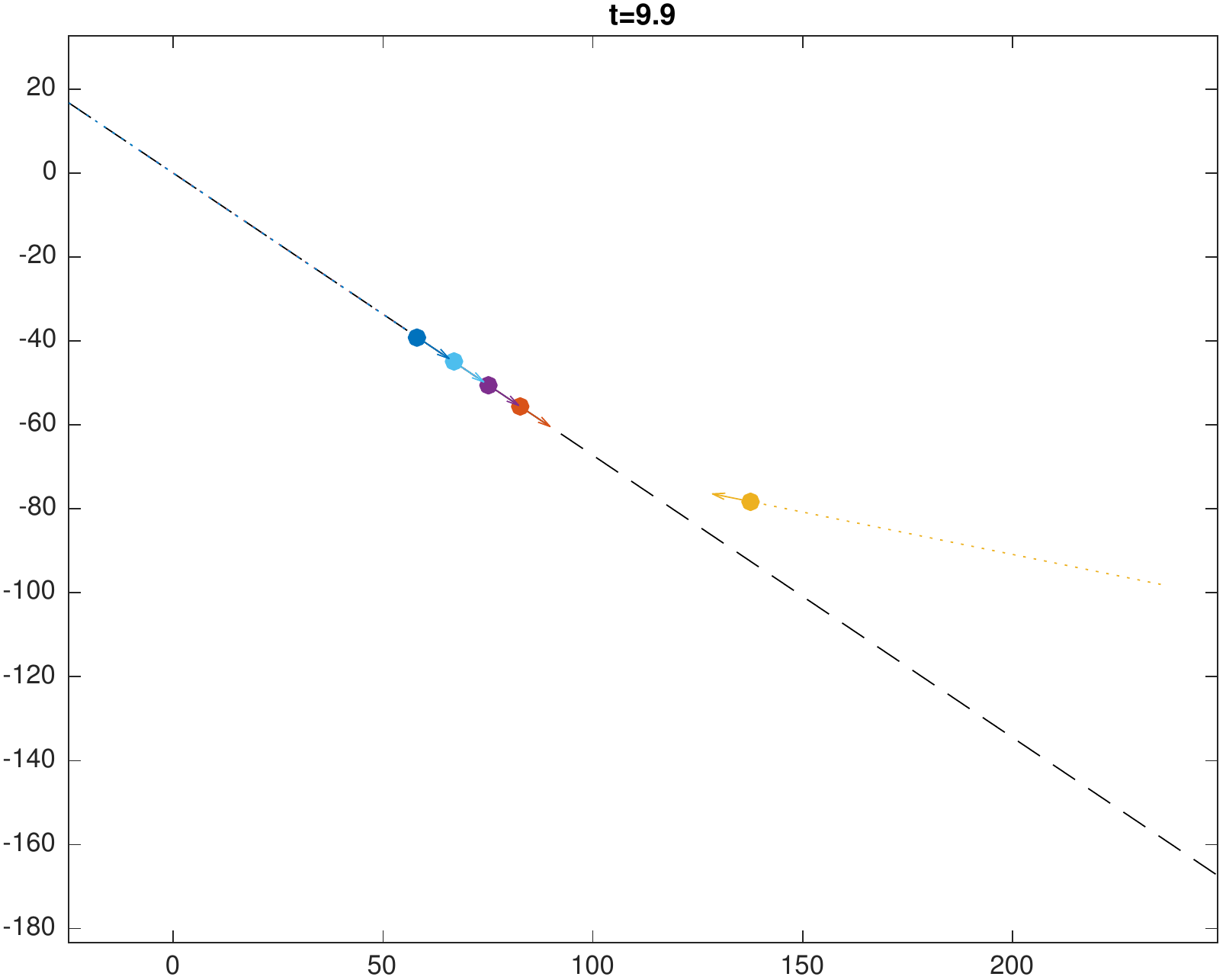}
        \caption{A four-vehicle platoon travels along the highway. The yellow vehicle disregards the others.}
    \end{subfigure}
    \begin{subfigure}[t]{0.9\columnwidth} \label{subfig:in_120}
        \includegraphics[width=\columnwidth]{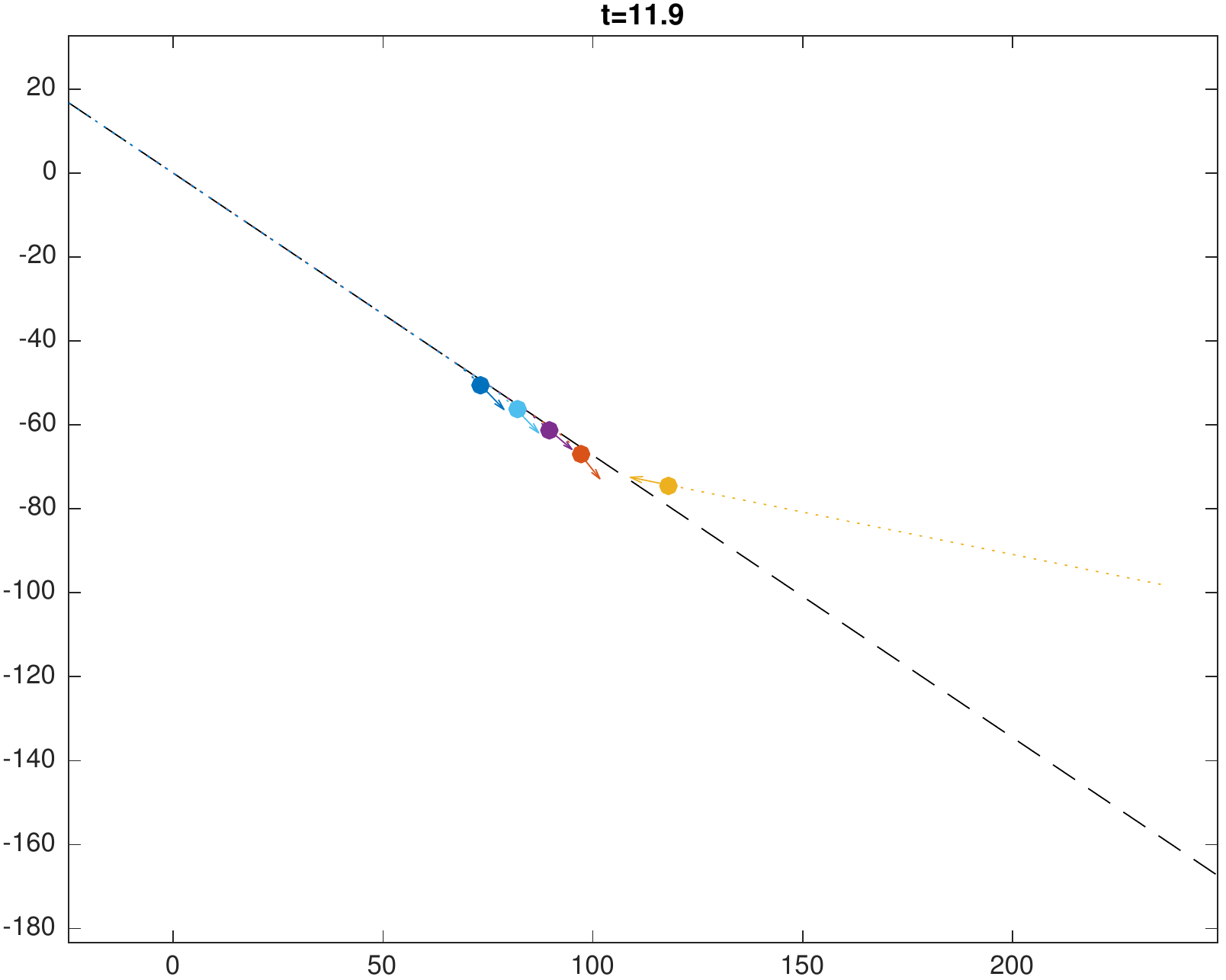}
        \caption{Vehicles begin avoidance maneuvers as they get near safety BRS boundaries.}
    \end{subfigure}

    \begin{subfigure}[t]{0.9\columnwidth} \label{subfig:in_140}
        \includegraphics[width=\columnwidth]{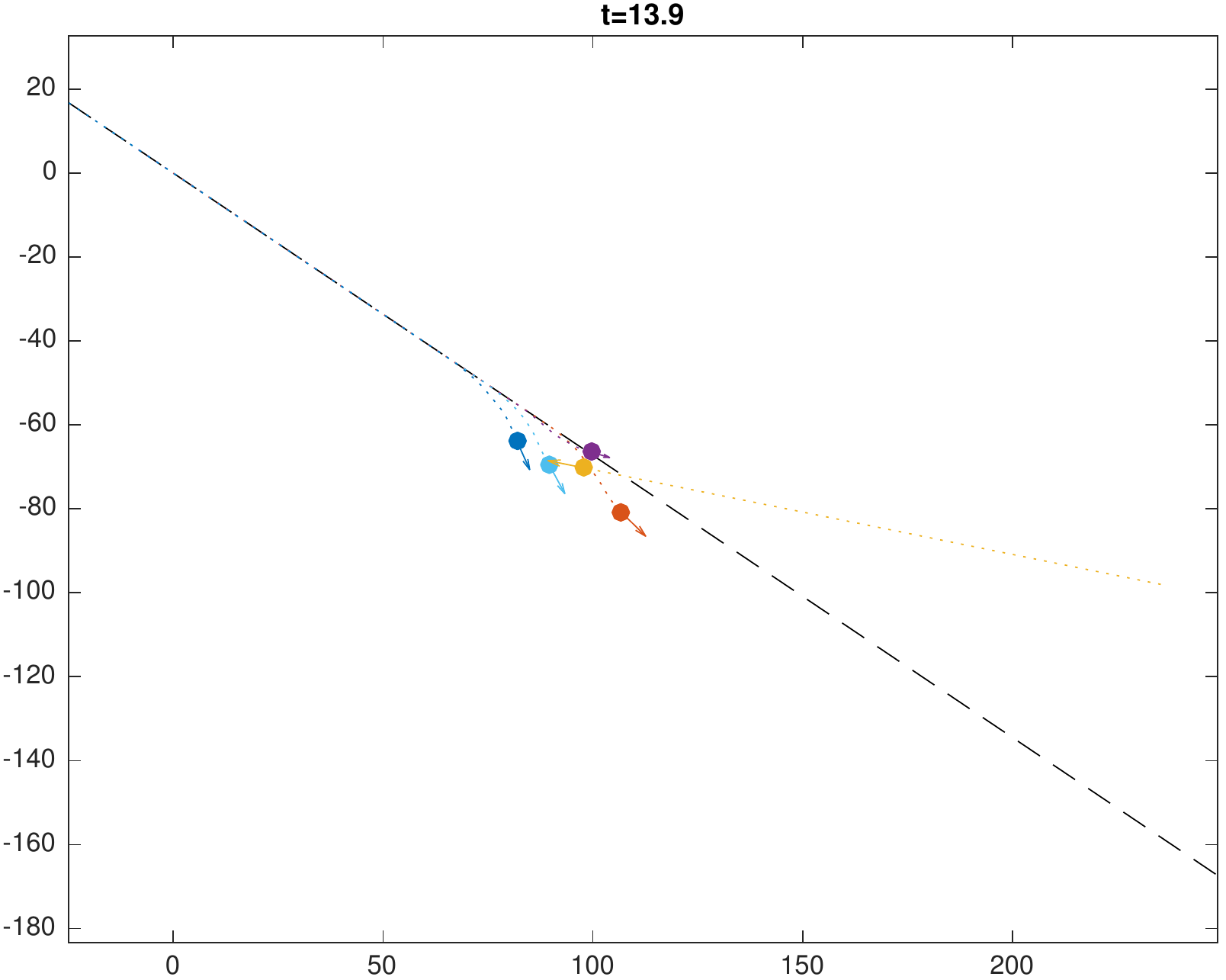}
        \caption{Safety controllers cause vehicles to spread out to ``make room'' for the intruder to pass.}
    \end{subfigure}
    \begin{subfigure}[t]{0.9\columnwidth} \label{subfig:in_200}
        \includegraphics[width=\columnwidth]{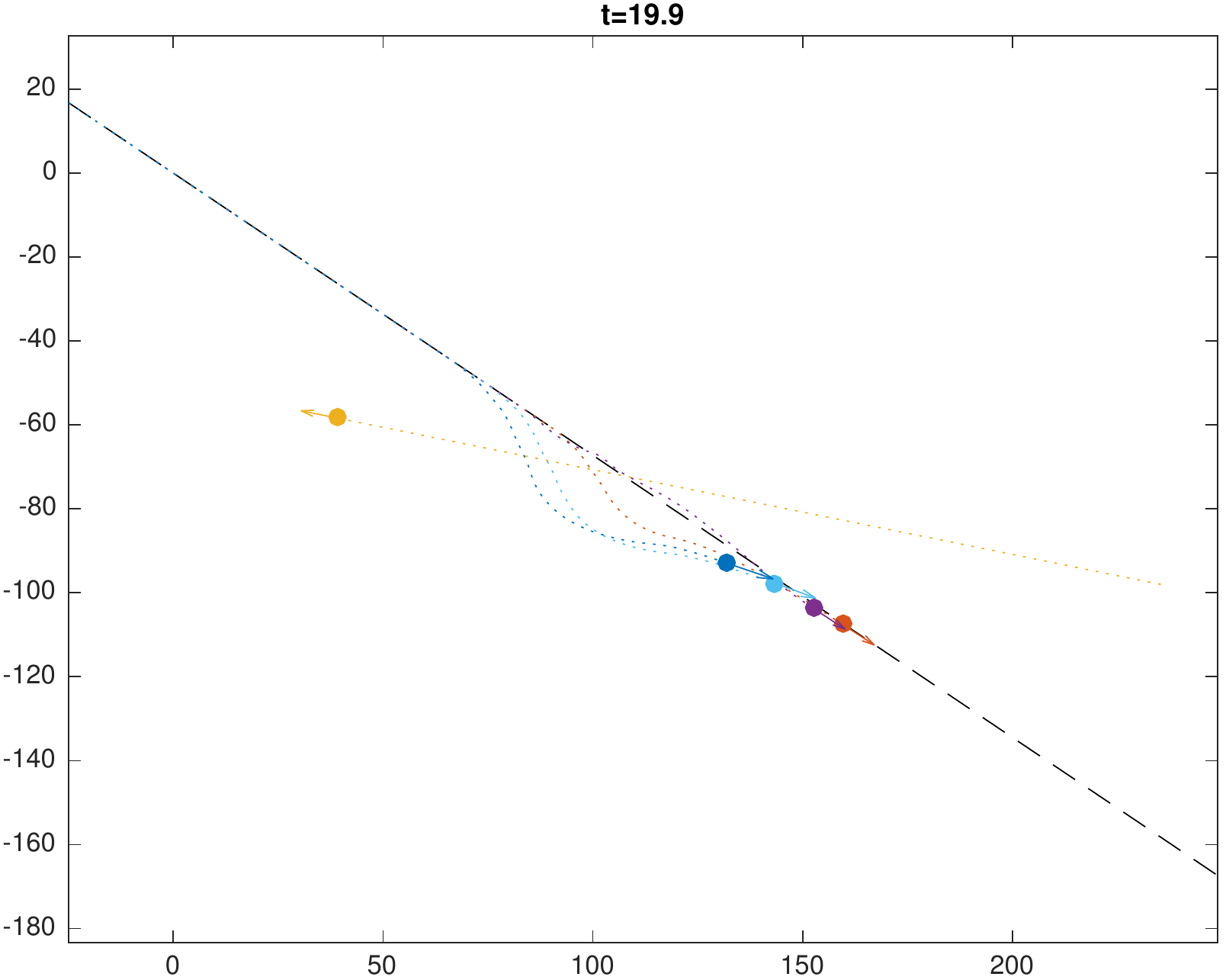}
        \caption{After danger has passed, the platoon resumes normal operation.}
    \end{subfigure}   
    \caption{A simulation showing how a platoon of four vehicles reacts to an intruder. \label{fig:in}}
\end{figure*}

Figure \ref{fig:in} shows the simulation result. At $t=9.9$, a platoon of four vehicles, $\veh{i},i=1,\ldots,4$ (with $P_i = i$), travels along the highway shown. An intruder vehicle $\veh{5}$ (yellow) heads left, disregarding the presence of the platoon. At $t=11.9$, the platoon leader $\veh{1}$ (red) detects that it has gone near the boundary of the safety BRS (not shown) with respect to the intruder $\veh{5}$ (yellow). In response, $\veh{1}$ (red) starts using the safety controller to optimally avoid the intruder according to \eqref{eq:HJI_ctrl_syn}; in doing so, it steers slightly off the highway. 

Note that although in this particular simulation, the intruder travels in a straight line, a straight line motion of the intruder was \textit{not} assumed. Rather, the safety BRSs are computed assuming the worst case control of the intruder, according to \eqref{eq:HJI_ctrl_syn}.

As the intruder $\veh{5}$ (yellow) continues to disregard other vehicles, the followers of the platoon also get near the respective boundaries of their safety BRSs with respect to the intruder. This occurs at $t=13.9$, where the platoon ``makes room'' for the intruder to pass by to avoid collisions; all vehicles deviate from their intended path, which is to follow the platoon leader or the highway. Note that in this case, we have assumed that the platoon does not move as a unit in response to an intruder to show more interesting behavior.

After the intruder has passed, eventually all vehicles become far away from any safety BRSs. When this occurs, the leader resumes following the highway, and the followers resume following the leader. At $t=19.9$, the platoon successfully gets back onto the highway.

\subsubsection{Changing highways}
\begin{figure*}[h!]
    \centering
    \begin{subfigure}[t]{0.9\columnwidth} \label{subfig:ch_83}
        \includegraphics[width=\columnwidth]{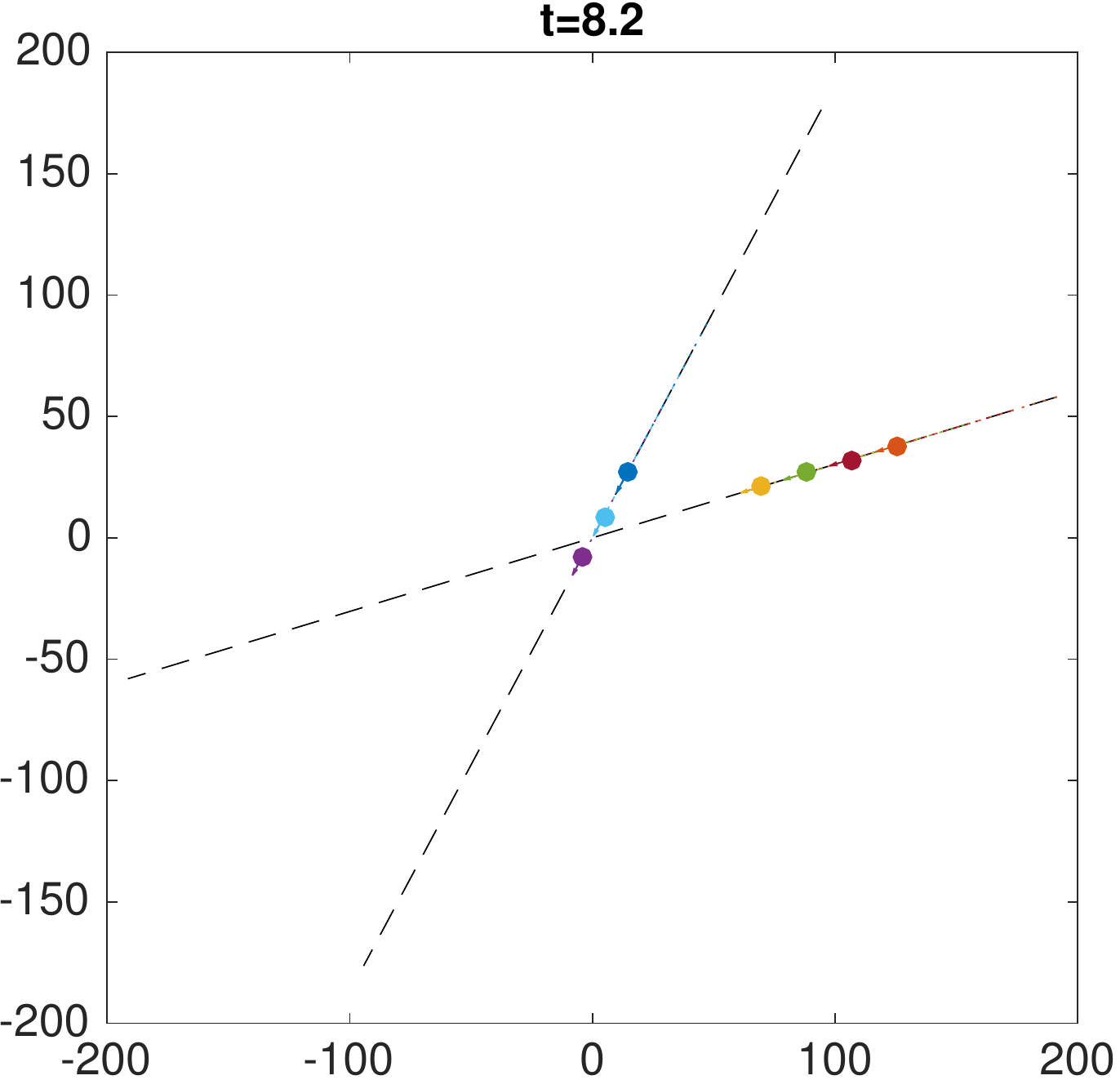}
        \caption{A three-vehicle platoon and a four-vehicle platoon travel on their respective air highways.}
    \end{subfigure}
    \begin{subfigure}[t]{0.9\columnwidth} \label{subfig:ch_124}
        \includegraphics[width=\columnwidth]{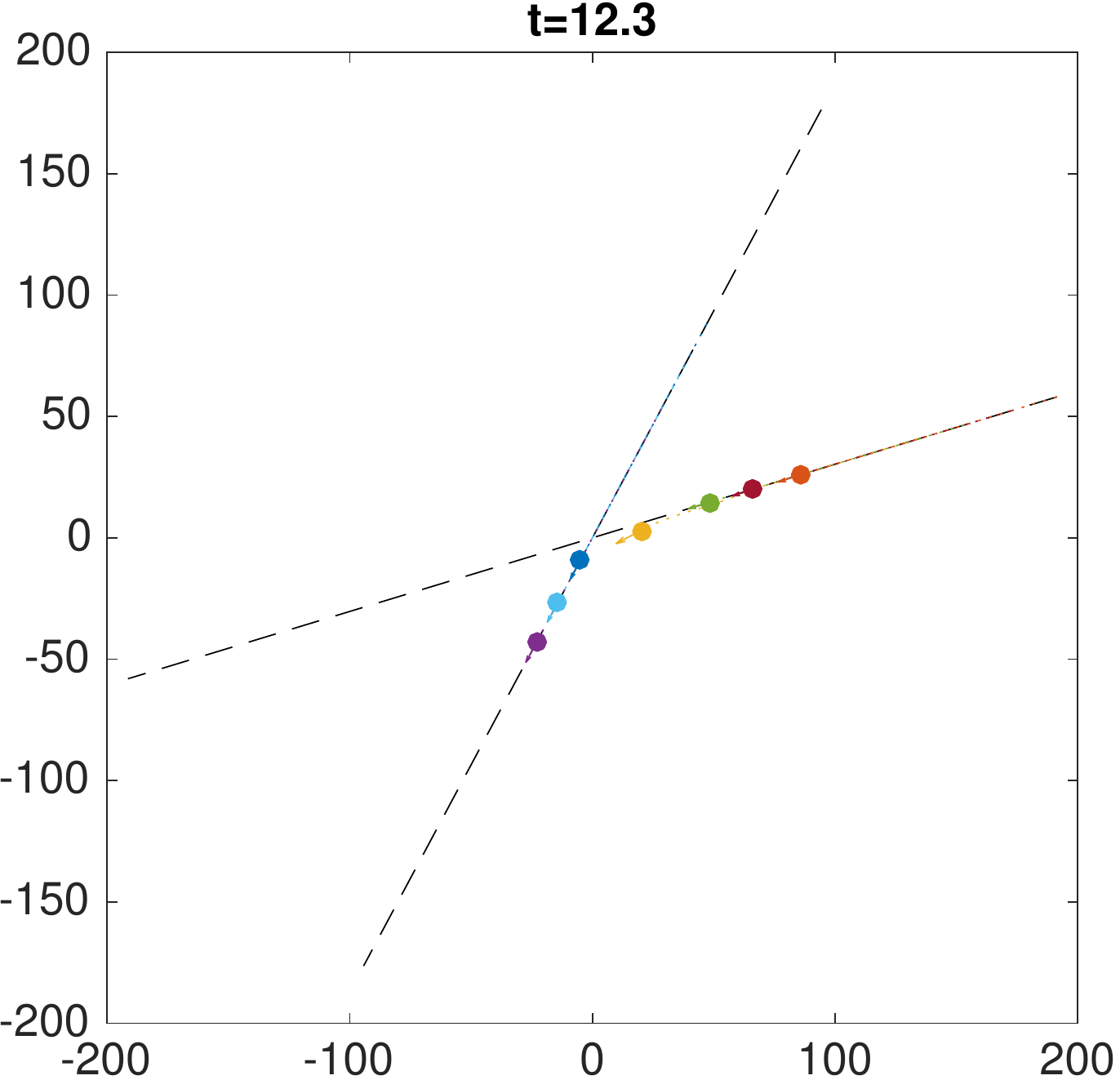}
        \caption{The yellow vehicle begins to join the new platoon; The green vehicle becomes a leader.}
    \end{subfigure}

    \begin{subfigure}[t]{0.9\columnwidth} \label{subfig:ch_170}
        \includegraphics[width=\columnwidth]{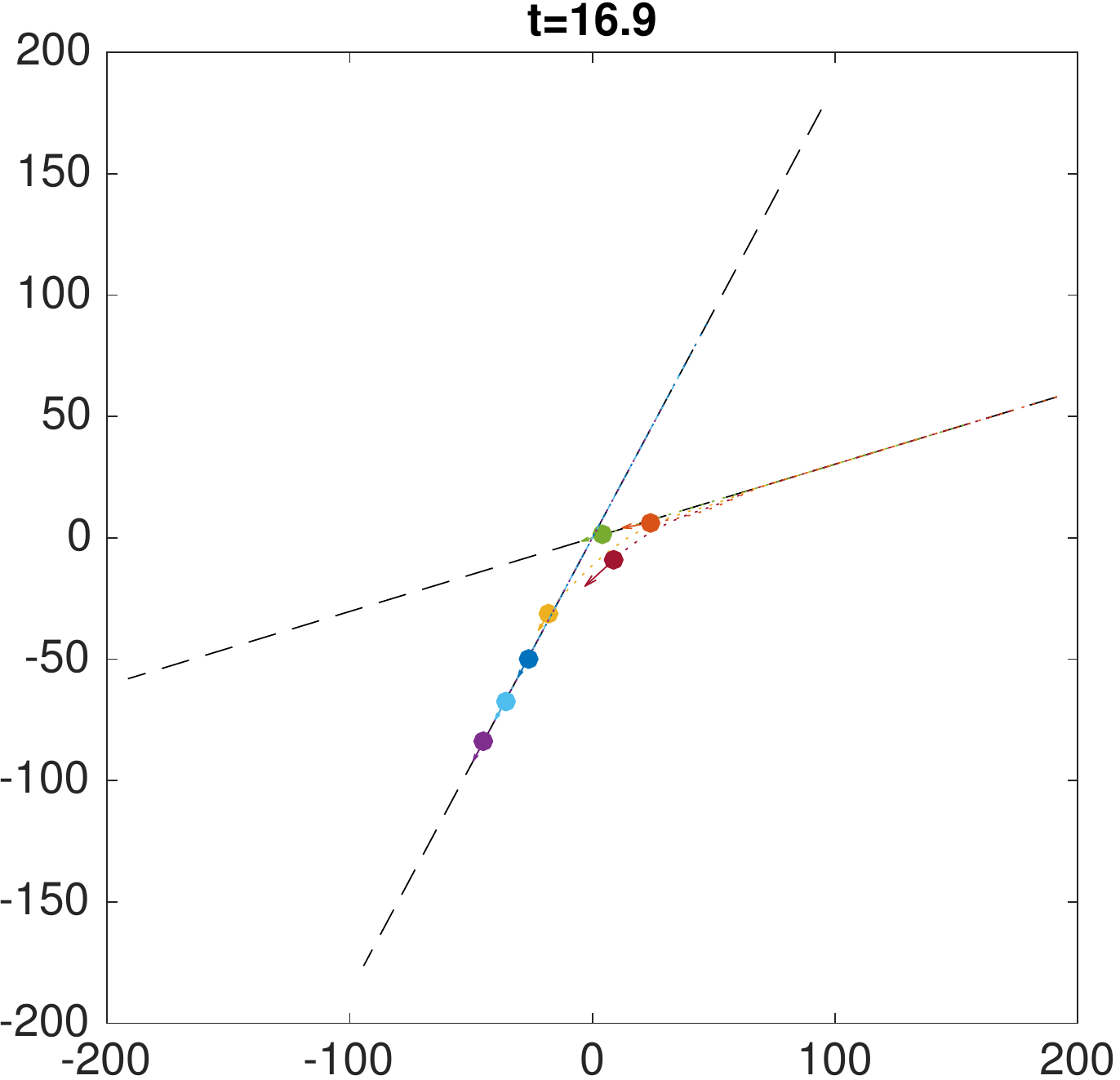}
        \caption{The dark red vehicle joins a new platoon; the orange vehicle catches up to new platoon leader.}
    \end{subfigure}
    \begin{subfigure}[t]{0.9\columnwidth} \label{subfig:ch_231}
        \includegraphics[width=\columnwidth]{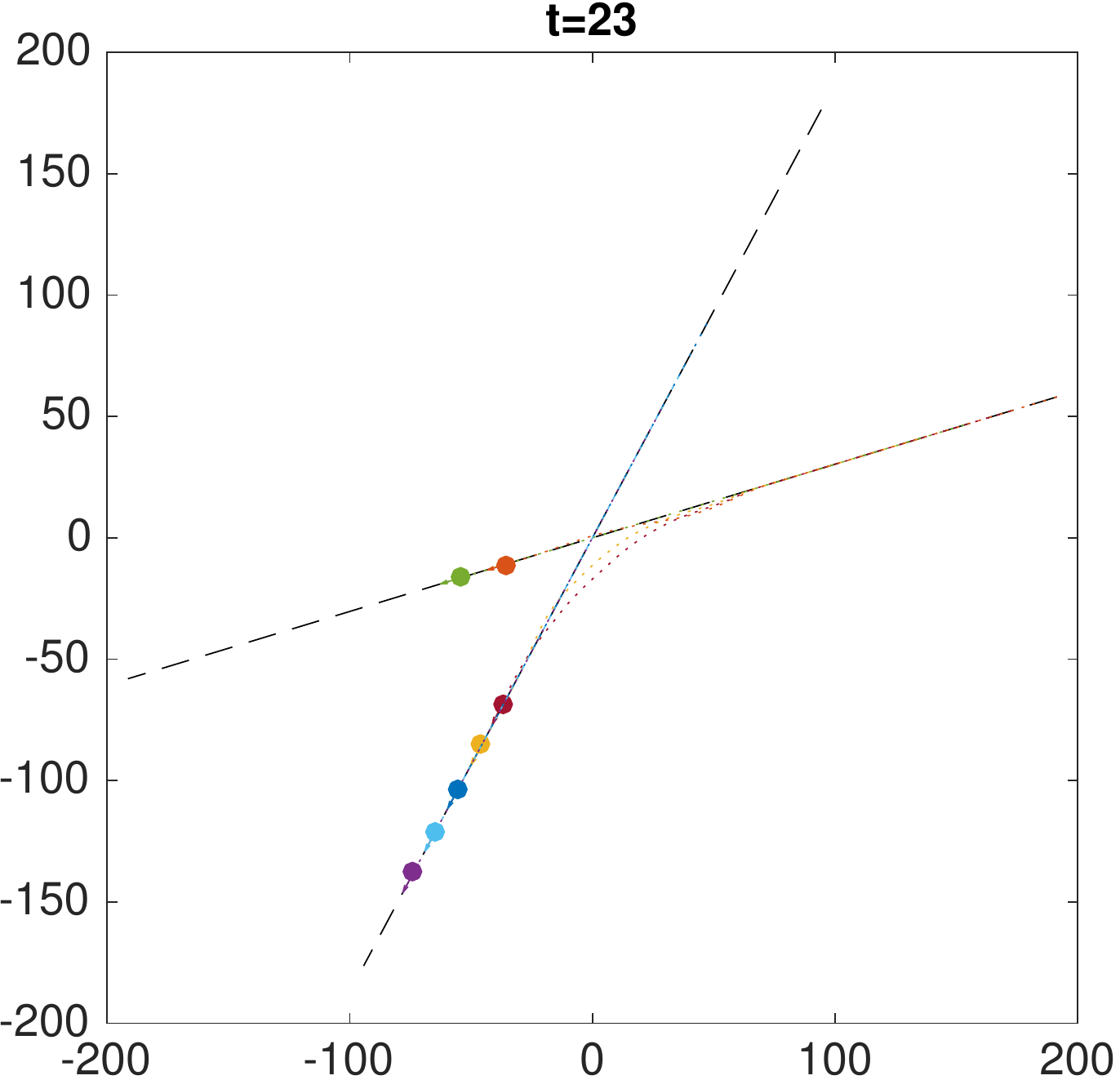}
        \caption{New platoons now travel on their respective air highways.}
    \end{subfigure}   
    \caption{A simulation showing two vehicles changing highways and joining a new platoon. \label{fig:ch}}
\end{figure*}

In order to travel from origin to destination, a vehicle may need to change highways several times before exiting an air highway system. In this simulation, shown in Figure \ref{fig:ch}, two platoons are traveling on two different highways that intersect. When the platoons are near the highway intersection, two of the vehicles in the four-vehicle platoon change highways and join the other platoon.

The $t=8.2$ plot shows the two platoons of vehicles traveling on the two air highways. One platoon has three vehicles, and the other has four vehicles. At $t=12.3$, the yellow vehicle begins steering off its original highway in order to join the other platoon. In terms of the hybrid systems modes, the yellow vehicle transitions from the Leader mode to the Follower mode. At the same time, the green vehicle transitions from the Follower mode to the Leader mode, since the previous platoon leader, the yellow vehicle, has left the platoon. By $t=16.9$, the yellow vehicle successfully changes highways and is now a follower in its new platoon.

At $t=16.9$, the dark red vehicle is in the process of changing highways. In this case, it remains in the Follower mode, since it is a follower in both its old and new platoons. While the dark red vehicle changes highways, the orange vehicle moves forward to catch up to its new platoon leader, the green vehicle. By $t=23$, all the vehicles have finished performing their desired maneuvers, resulting in a two-vehicle platoon and a five-vehicle platoon traveling on their respective highways.

\section{Conclusions}
To address the important and urgent problem of the traffic management of unmanned aerial vehicles (UAVs), we proposed to have platoons of UAVs traveling on air highways. We showed how such an airspace structure leads to much easier safety and goal satisfaction analysis. We provided simulations which show that by organizing vehicles into platoons, many complex maneuvers can be performed using just a few different backward reachable sets.

For the placement of air highways over a region, we utilize the very intuitive and efficient fast marching algorithm for solving the Eikonal equation. Our algorithm allows us to take as input any arbitrary cost map representing the desirability of flying over any position in space, and produce a set of paths from any destination to a particular origin. Simple heuristic clustering methods can then be used to convert the sets of paths into a set of air highways.

On the air highways, we considered platoons of UAVs modeled by hybrid systems. We show how various required platoon functions (merging onto an air highway, changing platoons, etc.) can be implemented using only the Free, Leader, and Follower modes of operation. Using HJ reachability, we proposed goal satisfaction controllers that guarantee the success of all mode transitions, and wrapped a safety controller around goal satisfaction controllers to ensure no collision between the UAVs can occur. Under the assumption that faulty vehicles can descend after a pre-specified duration, our safety controller guarantees that no collisions will occur in a single altitude level as long as at most one safety breach occurs for each vehicle in the platoon. Additional safety breaches can be handled by multiple altitude ranges in the airspace. 



\section*{Acknowledgments}
This work is supported in part by NSF under CPS:ActionWebs (CNS-0931843) and CPS:FORCES (CNS1239166), by NASA under grants NNX12AR18A and UCSCMCA-14-022 (UARC), by ONR under grants N00014-12-1-0609, N000141310341 (Embedded Humans MURI), and MIT\_5710002646 (SMARTS MURI), and by AFOSR under grants UPenn-FA9550-10-1-0567 (CHASE MURI) and the SURE project.

\bibliographystyle{aiaa}
\bibliography{references}
\end{document}